\documentclass[useAMS,usenatbib]{mn2e}
\usepackage{epsfig,rotate,graphicx}
\usepackage[fleqn]{amsmath}
\usepackage{subfigure}
\usepackage{lscape}
\usepackage{bm}

\newcommand{\p}{\partial}

\newcommand{\be}{\begin{equation}}
\newcommand{\ee}{\end{equation}}
\newcommand{\gtrsim}{\;\raisebox{-.8ex}{$\buildrel{\textstyle>}\over\sim$}\;}
\newcommand{\lesssim}{\; \raisebox{-.8ex}{$\buildrel{\textstyle<}\over\sim$}\;}

\newcommand{\sbar}{\bar{\sigma}}
\newcommand{\avg}[1]{\langle #1 \rangle}

\newcommand{\lmax}{l_\mathrm{max}}
\newcommand{\mmax}{m_\mathrm{max}}

\newcommand{\ii}{\mathrm{i}}
\newcommand{\bv}{\bm{v}}

\newcommand{\jlin}{j_\mathrm{lin}}
\newcommand{\jlintot}{J_\mathrm{lin}}
\newcommand{\tbg}{T_\mathrm{BG}}
\newcommand{\varA}[1]{{\operatorname{#1}}}

\DeclareMathOperator{\real}{Re}
\DeclareMathOperator{\imag}{Im}

\usepackage{array,booktabs,tabularx}
\newcolumntype{R}{>{\centering\arraybackslash}X} 

\title[One-armed spirals]{One-armed spirals in 
  locally isothermal, radially structured self-gravitating discs}     

\author[Lin]{Min-Kai Lin$^{1}$
  \thanks{ minkailin@email.arizona.edu} \\ 
  $^1$Department of Astronomy and Steward Observatory, University of
  Arizona, 933 North Cherry Avenue, Tucson, AZ 85721, USA 
}

\begin{document}

\maketitle
\begin{abstract} 
   We describe a new mechanism that leads to the destabilisation of
 non-axisymmetric waves in astrophysical discs with 
 an imposed radial temperature gradient. This might apply, for
 example, to the outer parts of protoplanetary discs. We use linear
 density wave theory 
 to show that non-axisymmetric perturbations generally do not conserve
 their angular momentum in the presence of a forced temperature
 gradient. This implies an exchange of angular momentum 
 between linear perturbations and the background disc. In particular, 
 when the disturbance is a low-frequency 
 trailing wave and the disc temperature decreases outwards, this 
 interaction is unstable and leads to the growth of the wave.  
 We demonstrate this phenomenon through numerical hydrodynamic
 simulations of locally isothermal discs in 2D using the FARGO
 code and in 3D with  the ZEUS-MP and PLUTO codes.  We consider
 radially structured discs with a self-gravitating region which remains
 stable in the absence of a temperature gradient. However,
 when a temperature gradient is imposed we observe exponential growth of a one-armed  
 spiral mode (azimuthal wavenumber $m=1$) with co-rotation radius 
 outside the bulk of the spiral arm, resulting in a nearly-stationary
 one-armed spiral pattern.  The development of this one-armed 
 spiral does not require the movement of the central star, as found in previous studies. 
 Because destabilisation by a forced temperature gradient does 
 not explicitly require disc self-gravity, we suggest this mechanism may also
 affect low-frequency one-armed oscillations in non-self-gravitating
 discs. 

\end{abstract}


\section{Introduction}\label{intro}
An exciting development in the study of circumstellar 
discs is the direct observation of large-scale, non-axisymmetric
structures within them. These
include lopsided dust distributions 
\citep{marel13,fukagawa13,casassus13,isella13,perez14,follette14,plas14} and
spiral arms 
\citep{hashimoto11,muto12,boccaletti14,grady13,christiaens14,avenhaus14}. 

The attractive explanation for asymmetries in circumstellar discs is
disc-planet interaction. In particular, spiral structures  
naturally arise from the gravitational interaction between a
planet and the gaseous protoplanetary disc it is embedded in
\citep[see, e.g.][for a recent review]{baruteau13b}. Thus, the
presence of spiral arms in circumstellar discs could be signposts of 
planet formation \citep[but see][]{juhasz14}.

Spiral arms are also  characteristic of global gravitational 
instability (GI)  in differentially rotating discs
\citep{goldreich65,laughlin96b,laughlin98,nelson98,lodato05,forgan11}. Large-scale 
spiral arms can provide significant angular momentum transport
necessary for mass accretion \citep{lynden-bell72,
  papaloizou91,balbus99,lodato04}, and spiral structures due to GI are
potentially observable with the Atacama Large
Millimeter/sub-millimeter Array \citep{cossins10,dipierror14}. GI can
be expected in the earliest stages of circumstellar disc formation 
\citep{kratter10b,inutsuka10,tsukamoto13}, and may be possible in the
outer parts of the disc \citep{rafikov05,matzner05,kimura12}.  

Single-arm spirals, or eccentric modes, corresponding to perturbations 
with azimuthal wavenumber $m=1$, have received interest in 
the context of protoplanetary discs because of their global nature
\citep{adams89,heemskerk92,laughlin96,tremaine01,papaloizou02,hopkins10}. 
In the `SLING' mechanism proposed by \cite{shu90}, an $m=1$ 
  gravitational instability arises from the motion of the
central star induced by the one-armed perturbation, and requires a
massive disc \citep[the former may have observable consequences,
][]{michael10}.

  In this work we identify a new mechanism that leads to the growth of
  one-armed spirals in astrophysical discs. We show
  that when the disc temperature is prescribed (called locally isothermal
  discs), the usual statement of the conservation of angular momentum
  for linear perturbations acquires a source term proportional to the
  temperature gradient. This permits angular momentum exchange between
  linear perturbations and the background disc. This `background
  torque' can destabilise low-frequency non-axisymmetric
  trailing waves when the disc temperature decreases outwards.

 We employ direct hydrodynamic simulations using three different
  grid-based codes to demonstrate how this `background torque' 
  can lead to the growth of one-armed spirals in radially structured, 
  self-gravitating discs. This is despite the fact that our disc
  models do not meet the requirements for the `SLING' 
  mechanism. Although our numerical simulations consider
  self-gravitating discs, this `background torque' is generic for
  locally isothermal discs and its existence does not require
  disc self-gravity. Thus, the destabilisation effect we
  describe should also be applicable to non-self-gravitating discs.

This paper is organised as follows. In \S\ref {model} we describe the system of interest
and list the governing equations. In \S\ref{wkb} we use linear theory  to show
  how a fixed temperature profile 
  can destabilise non-axisymmetric waves in discs. \S\ref{methods} describes
the numerical setup and hydrodynamic codes we use to 
  demonstrate the growth of one-armed spirals due to an imposed
  temperature gradient. Our
  simulation results are presented in  \S\ref{results2d} for two-dimensional (2D) 
  discs and in \S\ref{results3d} for three-dimensional (3D) discs, and we further
discuss them in \S\ref{discussions}. We summarise in \S\ref{summary}  
with some speculations for future work. 

\section{Governing equations}\label{model} 
We consider an inviscid fluid disc of mass $M_d$ 
orbiting a central star of mass $M_*$. We will mainly examine 2D
(or razor-thin) discs in favour of numerical 
resolution, but have also carried out some 3D disc
simulations. Hence, for generality we describe the system in
3D, using both cylindrical $(R,\phi,z)$ and spherical polar
coordinates $(r,\theta,\phi)$ centred on the star. 
The governing fluid equations in 3D are  
\begin{align}
  &\frac{\p\rho}{\p t} + \nabla\cdot\left(\rho\bm{v}\right) =
  0,\label{cont_eq}\\
  &\frac{\p\bm{v}}{\p t} + \bm{v}\cdot\nabla\bv = -\frac{1}{\rho}\nabla
  p -\nabla \Phi_\mathrm{tot}\label{mom_eq},\\ 
  & \nabla^2\Phi_d = 4 \pi G \rho \label{poisson}, 
\end{align}
where $\rho$ is the mass density, $\bm{v}$ is the 
fluid velocity (the angular velocity being $\Omega \equiv v_\phi/R$), 
$p$ is the pressure and the total potential
$\Phi_\mathrm{tot} = \Phi_* + \Phi_d$ consists of that from the
central star, 
\begin{align}
  \Phi_*(r) = -\frac{GM_*}{r}, 
\end{align}
where $G$ is the gravitational constant,  and the disc potential
$\Phi_d$. We impose a locally isothermal equation of state 
\begin{align}
  p = c_s^2(R)\rho,
\end{align}
where the sound-speed $c_s$ is given by 
\begin{align}\label{sound-speed}
  c_s(R) = c_{s0}\left(\frac{R}{R_0}\right)^{-q/2}
\end{align}
where $c_{s0} = h R_0\Omega_k(R_0)$ and 
$h$ is the disc aspect-ratio at the reference radius $R=R_0$, 
$\Omega_k=\sqrt{GM_*/R^3}$ is the midplane Keplerian frequency, and
$-q$ is the imposed temperature gradient since, for an ideal gas the
temperature is proportional to $c_s^2$. For convenience we will
  refer to $c_s^2$ as the disc temperature. The vertical disc scale-height is
defined by $H=c_s/\Omega_k$. Thus, a strictly isothermal disc with
$q=0$ has $H\propto R^{3/2}$, and $q=1$ corresponds to a 
 disc with constant $H/R$.

The 2D disc equations are obtained by replacing $\rho$ with the surface
mass density $\Sigma$, $p$ becomes the vertically-integrated pressure, 
and the 2D fluid velocity $\bm{v}$ is evaluated at the midplane, as are
the forces in the momentum equations. In the Poisson equation, $\rho$ is
replaced by $\Sigma\delta(z)$, where $\delta(z)$ is the
delta function.

\section{Linear density waves}\label{wkb}

 We describe a key feature of locally isothermal discs that
  enables angular momentum exchange between small disturbances and the
  background disc through an imposed radial temperature gradient. This
  conclusion results from the consideration of angular momentum
  conservation within the framework of linear perturbation theory. 
  For simplicity, in this section we consider 2D discs.

In a linear analysis, one assumes a steady axisymmetric background state, 
which is then perturbed such that
\begin{align}  
  \Sigma \to \Sigma(R) + \delta\Sigma_m(R)\exp{\left[\ii\left(-\sigma t +
        m\phi\right)\right]}, 
\end{align}
and similarly for other variables, where $\sigma=\omega+\ii\gamma$ is
a complex frequency with $\omega$ being the real frequency, 
$\gamma$ the growth rate, and $m$ is an integer. We take $m>0$ without
loss of generality.

The linearised mass and momentum equations are
\begin{align}
  &-\ii\sbar \delta\Sigma_m = -\frac{1}{R}\frac{d}{dR}\left(R\Sigma\delta
    v_{Rm}\right) - \frac{\ii m \Sigma}{R}\delta v_{\phi m}, \\
  &-\ii\sbar\delta v_{Rm} - 2\Omega \delta v_{\phi m} = -
  c_s^2(R)\frac{d}{dR}\left(\frac{\delta\Sigma_m}{\Sigma}\right) - \frac{d}{dR}\delta\Phi_m,\label{radmom}\\
  & -\ii\sbar\delta v_{\phi m} + \frac{\kappa^2}{2\Omega}\delta v_{Rm} =
  -\frac{\ii m }{R}\left(c_s^2\frac{\delta\Sigma_m}{\Sigma} + \delta\Phi_m\right),
\end{align}
where $\sbar = \sigma - m\Omega$ and $\kappa^2 =
R^{-3}\p_R(R^4\Omega^2)$ is the square of the epicyclic frequency. A
locally isothermal equation of state has been assumed in
Eq. \ref{radmom}.  The linearised Poisson equation is 
\begin{align}
  \nabla^2\delta\Phi_m = 4\pi G \delta\Sigma_m \delta(z). 
\end{align}

These linearised equations can be combined into a single
integro-differential equation eigenvalue problem. We defer a full
 numerical exploration of the linear problem to a future study. Here, we 
  discuss some
general properties of the linear perturbations. 

\subsection{Global angular momentum conservation for linear
  perturbations} \label{global_cons}
It can be shown that linear perturbations with $\phi$-dependence in the form
$\exp{(\ii m\phi)}$ satisfies angular momentum conservation
in the form 
\begin{align}\label{angcons}
  \frac{\p \jlin}{\p t} + \nabla\cdot\bm{F} = \tbg, 
\end{align}
\citep[e.g.][]{narayan87,ryu92,lin93b} where
\begin{align}\label{lin_ang_mom_cons}
  \jlin \equiv
  -\frac{m\Sigma}{2}\imag\left(\bm{\xi}^*\cdot\frac{\p\bm{\xi}}{\p
      t} + \Omega \hat{\bm{k}}\cdot\bm{\xi}\times\bm{\xi}^* + \ii
    m \Omega |\bm{\xi}|^2\right) 
\end{align}
is the angular momentum density of the linear disturbance (which may
be positive or negative), $\bm{\xi}$ is the Lagrangian
displacement and $^*$ denotes complex conjugation, and $\bm{F}$ is the
vertically-integrated angular momentum flux consisting of a Reynolds
stress and a gravitational stress \citep{lin11b}. Explicit expressions
for $\bm{\xi}$ can be found in, e.g. \cite{papaloizou85}. 

In Eq. \ref{angcons}, the background torque density $\tbg$ is 
\begin{align}\label{baroclinic_torque}
  \tbg \equiv
  -\frac{m}{2}\imag\left(\delta\Sigma_m\xi_R^*\frac{dc_s^2}{dR}\right), 
\end{align}
and arises because we have adopted a locally isothermal equation of
state in the perturbed disc.  We outline the
derivation of $T_\mathrm{BG}$ in Appendix \ref{tbg_deriv}. 

In a barotropic fluid, such as a strictly isothermal disc,
$\tbg$ vanishes and the total angular momentum associated with
the perturbation is conserved, provided that there is no net angular
momentum flux. However, as noted in \cite{lin11b}, if there is an imposed
temperature gradient, as in the disc models we consider,
then $\tbg\neq 0$ in general, which corresponds to a local torque
exerted by the background disc on the perturbation. 

The important consequence of the background torque is the possibility
of instability if $\tbg\jlin>0$. That is, if $\jlin$ is positive
(negative) and $\tbg$ is also positive (negative),
then the local angular momentum density of the linear disturbance
 will further increase (decrease) with time. This implies the amplitude of the disturbance
may grow by exchanging angular momentum with the background
disc.

  We demonstrate instability for low-frequency modes 
  ($|\omega|\ll m\Omega$) by explicitly showing its 
  angular momentum density 
  $\jlin<0$, and the background torque $T_\mathrm{BG}<0$ for
  appropriate perturbations and radial temperature gradients.

\subsection{Angular momentum density of  non-axisymmetric low-frequency modes}
From Eq. \ref{lin_ang_mom_cons} and assuming a time-dependence of the
form $\exp{(-\ii \sigma t)}$ with $\real{\sigma} = \omega$,  
the angular momentum density associated with linear waves is
\begin{align}
  \jlin = \frac{m\Sigma}{2}\left[\left(\omega -
      m\Omega\right)|\bm{\xi}|^2 + \ii\Omega\left(\xi_R\xi_\phi^* -
      \xi_R^*\xi_\phi\right)\right].\label{ang_mom_def}  
\end{align}
For a low-frequency mode, $|\omega|\ll m\Omega$. Then neglecting the
term $\omega|\bm{\xi}|^2$,  we find 
\begin{align}
  \jlin &\simeq \frac{m\Sigma\Omega}{2}\left[-m|\bm{\xi}|^2 + \ii\left(\xi_R\xi_\phi^* -
      \xi_R^*\xi_\phi\right)\right]\notag\\
  & = -\frac{m\Sigma\Omega}{2}\left[ (m-1)|\bm{\xi}|^2 +  |\xi_R + \ii\xi_\phi|^2\right].
\end{align}

Thus, non-axisymmetric ($m\geq1$) low-frequency modes generally have 
negative angular momentum. If the mode loses (positive) angular momentum 
to the background, then we can expect instability. We show 
below how this is possible through a forced temperature
gradient via the background torque. It is simplest to calculate
$T_\mathrm{BG}$ in the local approximation, which we review first.


\subsection{Local results}\label{local_approx}
In the local approximation, perturbations are assumed to vary rapidly
relative to any background gradients. The
dispersion relation for tightly-wound density 
waves of the form $\exp{[\ii(-\sigma t + m \phi + kR)]}$ in a razor-thin
disc is  
\begin{align}\label{dispersion}
  (\sigma - m\Omega)^2 = \kappa^2 + k^2c_s^2 - 2\pi G \Sigma |k|, 
\end{align}
where $k$ is a real wavenumber such that $|kR|\gg1$ \citep{shu91}. 
Note that in the strictly local approximation, where all
  global effects are neglected, only axisymmetric perturbations
($m=0$) can be unstable. 

Given the real frequency $\omega$ or pattern speed $\Omega_p\equiv  
\omega/m$ of a non-axisymmetric neutral mode 
, Eq. \ref{dispersion} can be solved 
for $|k|$, 
\begin{align}\label{wavenumber}
|k| = k_c\left[1 \pm \sqrt{1 -
     Q^2(1-\nu^2)}\right], 
\end{align}
where 
 \begin{align}
   k_c \equiv \frac{\pi G \Sigma}{c_s^2}
 \end{align}
 is a characteristic wavenumber, 
\begin{align}
  Q \equiv \frac{c_s\kappa}{\pi G \Sigma}
\end{align}
is the usual Toomre parameter, and
\begin{align}
  \nu \equiv \frac{(\omega - m\Omega)}{\kappa}
\end{align}
is a dimensionless frequency. In
Eq. \ref{wavenumber}, the upper (lower) sign correspond to short
(long) waves, and $k>0$ ($k<0$) correspond to trailing (leading)
waves.    

 At the \emph{co-rotation radius} $R_c$ the pattern speed matches
the fluid rotation,
\begin{align}
  \Omega(R_c) = \Omega_p.
\end{align}
\emph{Lindblad resonances} $R_L$ occurs where
\begin{align}
  \nu^2(R_L) = 1. 
\end{align}
Finally, \emph{Q-barriers} occur at radii $R_{Qb}$ where
\begin{align}
  Q^2(R_{Qb})\left[1-\nu^2(R_{Qb})\right] = 1.  
\end{align}
According to Eq. \ref{wavenumber}, purely wave-like solutions with
real $k$ are only possible where $Q^2(1-\nu^2)\leq1$.  

A detailed discussion of the properties of local density waves 
is given in \cite{shu91}. An important result, which holds for
  waves of all frequencies, is that
waves interior to co-rotation ($R<R_c$) have negative angular momentum, while
waves outside co-rotation ($R>R_c$) have positive angular
momentum. 

\subsection{Unstable interaction between low-frequency modes 
  and the background disc due to an imposed temperature gradient}
 
Here we show that the torque density acting on a 
local mode due to the background temperature gradient can be negative, which 
would enforce low-frequency modes, because they have negative angular momentum.

The Eulerian surface density perturbation is given by
\begin{align}\label{den_pert}
  \delta\Sigma_m = -\nabla\cdot\left(\Sigma\bm{\xi}\right) 
  = -\frac{1}{R}\frac{d}{dR}\left(R\Sigma \xi_R\right) - \frac{\ii m}{R}\Sigma\xi_\phi.  
\end{align}
We invoke local theory by setting $d/dR \to \ii k$ where $k$ is
real, and assume $|kR|\gg m$ so that the second term on the right hand
side of Eq. \ref{den_pert} can be neglected. Then    
\begin{align}
  \delta\Sigma _m  \simeq -\ii k \Sigma \xi_R.
\end{align}
 Inserting this into the expression for the background torque,
  Eq. \ref{baroclinic_torque}, we find 
\begin{align}
  T_\mathrm{BG} = \frac{m}{2}\frac{dc_s^2}{dR}k\Sigma |\xi_R|^2. \label{baroclinic_torque1}
\end{align}

This torque density is negative for trailing waves ($k>0$) in discs with
a fixed temperature profile decreasing outwards ($dc_s^2/dR<0$). Note
that this conclusion does not rely on the low-frequency approximation.    

However, if the linear disturbance under consideration \emph{is} a
low-frequency mode, then it has negative angular
momentum. If it is trailing and $dc_s^2/dR<0$, as is typical
in astrophysical discs, then $T_\mathrm{BG}<0$ and
the background disc applies a negative torque on the disturbance,
which further decreases its angular momentum. This suggests the 
  mode  
amplitude will grow.  

Using $\jlin$ and $T_\mathrm{BG}$, we can estimate the 
growth rate $\gamma$ of linear perturbations due to the background
torque as  
\begin{align}
  2\gamma \sim \frac{T_\mathrm{BG}}{\jlin},
\end{align}
where the factor of two accounts for the fact that the angular momentum
density is quadratic in the linear perturbations. Inserting the above
expressions for $\jlin$ and $T_\mathrm{BG}$ for gives
\begin{align}\label{theoretical_rate0}
  2\gamma \sim
  -\frac{dc_s^2}{dR}
  \frac{k}{\Omega}\frac{|\xi_R|^2}{\left[(m-1)|\bm{\xi}|^2 + |\xi_R +
      \ii\xi_\phi|^2 \right]} 
  = -\frac{dc_s^2}{dR}
  \frac{k}{m\Omega},
\end{align}
 where the second equality uses $\xi_\phi \simeq
2\ii \xi_R/m$ for low-frequency modes in the local approximation, as shown in  Appendix
\ref{horizontal_displacements}. 
Then for the temperature profiles $c_s^2 = c_{s0}^2 (R/R_0)^{-q}$ as
adopted in our disc models,  
\begin{align}
  2\gamma \sim q\frac{c_s^2}{R}\frac{k}{m\Omega} \sim q h
  \left(\frac{kH}{m}\right)\Omega,\label{theoretical_rate}
\end{align}
where we used $c_s\sim H\Omega\sim hR\Omega$. 
  Eq. \ref{theoretical_rate} suggests that perturbations with small
  radial length-scales ($kH\gtrsim m$) are most favourable for
  destabilisation. Taking the local approximation is then
  appropriate.

 Note that the derivation of Eq. \ref{baroclinic_torque1} and 
  Eq. \ref{theoretical_rate} do not 
  require the disc to be self-gravitating. Thus, destabilisation by the
  background torque is not directly associated with disc
  self-gravity. However, in order to evaluate 
  Eq. \ref{baroclinic_torque1} or Eq. \ref{theoretical_rate} in terms 
  of disc parameters (as done in \S\ref{fargo_m1}), we need to insert a 
  value of $k$, which may depend on disc    
  self-gravity (e.g. from Eq. \ref{wavenumber}).

\section{Numerical simulations}\label{methods}

  We demonstrate the destabilising effect of a fixed temperature
  gradient using numerical simulations. The above discussion is
  generic for low-frequency non-axisymmetric modes, but for
  simulations we will consider specific examples.  
  
  There are two parts to the destabilising mechanism: 
  the disc should support low-frequency modes, which is then
  destabilised by an imposed temperature gradient. The latter is
  straight forward to implement by adopting a locally isothermal
  equation of state as described in \S\ref{model}.  For the former, we
  consider discs with a  
  radially-structured Toomre $Q$ profile. We use local theory to show
  that such discs can trap 
  low-frequency one-armed ($m=1$) modes.  
  This is convenient because Eq. \ref{theoretical_rate} indicates that
  modes with small $m$ are
  more favourable for destabilisation.

\subsection{Disc model and initial conditions}
For the initial disc profile we adopt a modified power-law disc with
surface density given by
\begin{align}
  \Sigma(R) = \Sigma_\mathrm{ref} \left(\frac{R}{R_0}\right)^{-s}\times B(R;
  R_{1}, R_{2}, \epsilon, \delta R), 
\end{align}
where $s$ is the power-law index describing the smooth disc and
$\Sigma_\mathrm{ref}$ is a 
surface density scale chosen by specifying $Q_\mathrm{out}$,
the Keplerian Toomre parameter at $R=R_{2}$,
\begin{align}
  Q_\mathrm{out} = \left.\frac{c_s\Omega_k}{\pi G
    \Sigma}\right|_{R=R_{2}}. 
\end{align}
The bump function
$B(R)$ represents a surface density boost between
$R\in[R_{1},R_{2}]$ by a factor $\epsilon^{-1}>1$,
and $\delta R$ is the transition width between the bump and the
smooth disc. We choose 
\begin{align}\label{sig_bump}
  &B(R) = f_1(R)\times f_2(R),\\
  &f_1(R) = \frac{1}{2}\left(1 - \epsilon\right)\left[1 +
    \tanh\left(\frac{R-R_{1}}{\Delta_1}\right)\right]  + \epsilon,\\
  &f_2(R) = \frac{1}{2}\left(1 - \epsilon\right)\left[1 -
    \tanh\left(\frac{R-R_{2}}{\Delta_2}\right)\right]  + \epsilon,
\end{align}
where $\Delta_{1,2} = \delta R \times H(R_{1,2})$. 

The 3D disc structure is obtained by assuming vertical hydrostatic
balance  
\begin{align}
  0 = \frac{1}{\rho}\frac{\p p}{\p z} + \frac{\p\Phi_*}{\p z} + \frac{\p
    \Phi_d}{\p z},  
\end{align}
which gives the mass density as 
\begin{align}
  \rho = \frac{\Sigma}{\sqrt{2\pi}H}Z(R,z),
\end{align}
where $Z(R,z)$ describes vertical stratification. In practice, we
numerically solve for $Z(R,z)$ by neglecting the radial self-gravity
force compared to vertical self-gravity, which reduces the equations
for vertical hydrostatic equilibrium to ordinary differential
equations. This procedure is described in \cite{lin12b}. 

\begin{figure}
  \includegraphics[width=\linewidth,clip=true,trim=0cm 1.7cm 0cm
  0cm]{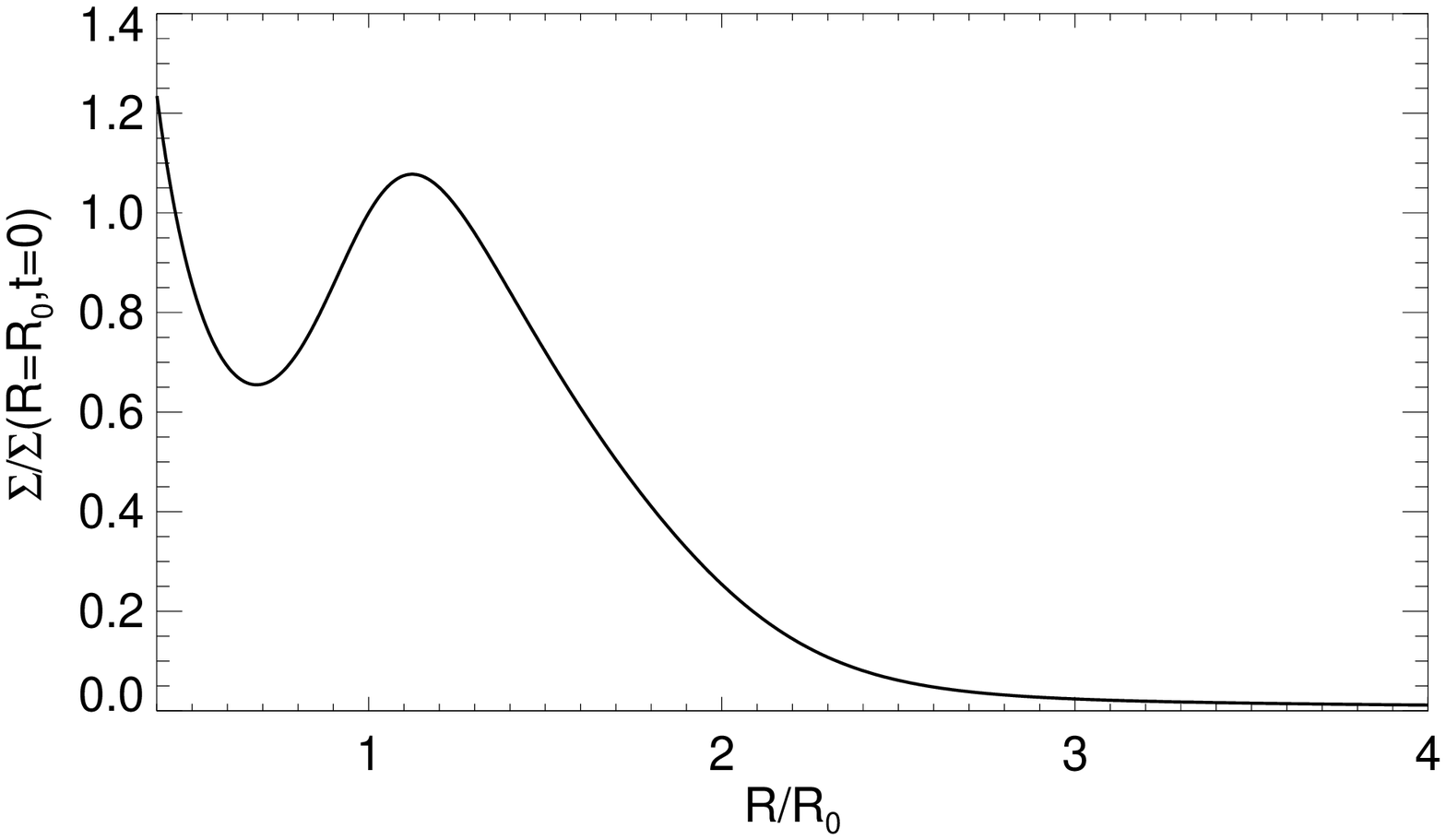} 
  \includegraphics[width=\linewidth]{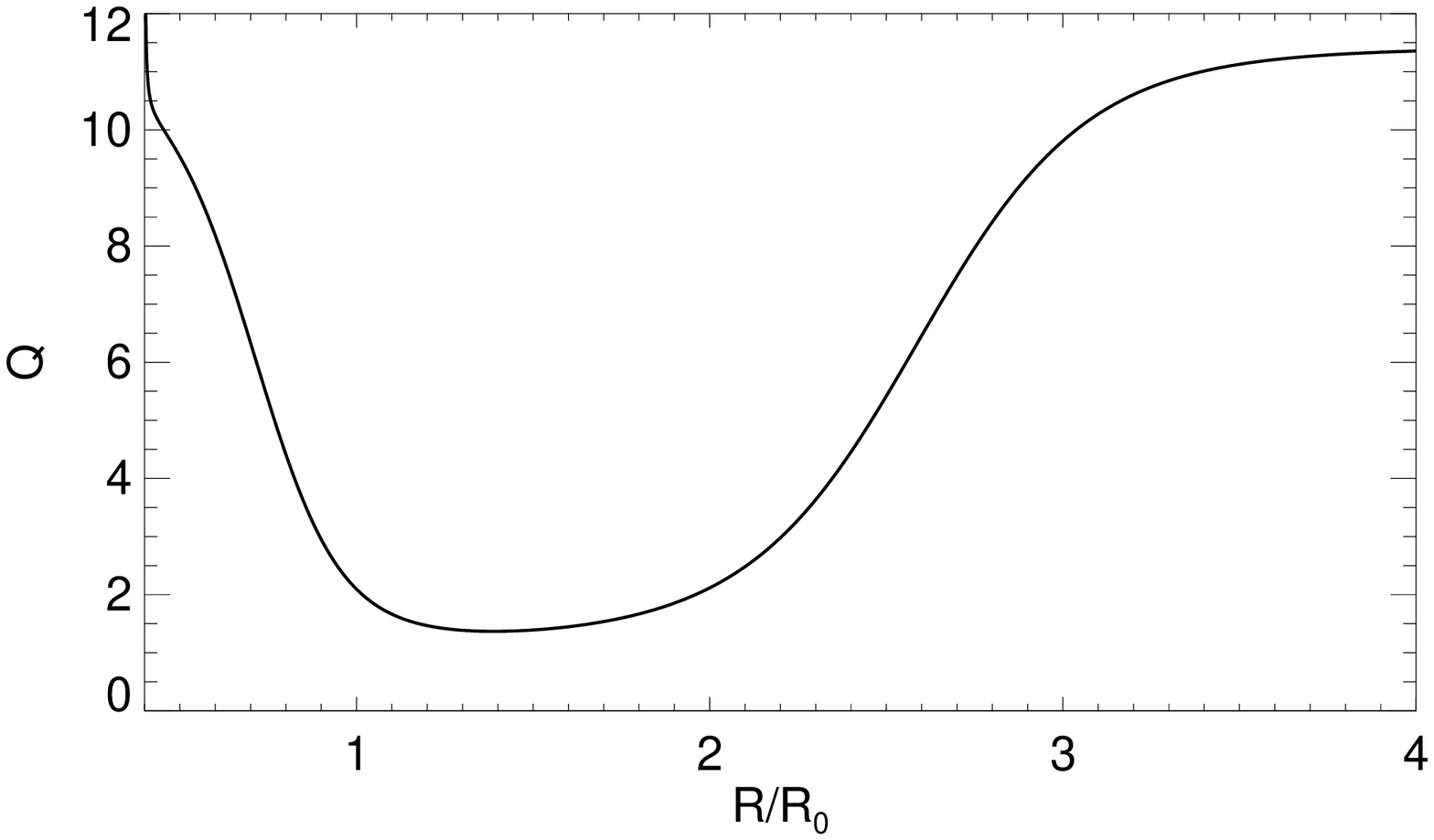}
  \caption{Fiducial profiles of the surface density (top) and Toomre
    parameter (bottom) used in this work.\label{initial_surf}}
\end{figure}

Our fiducial parameter values are: $s=2$, $R_{1}=R_0$, $R_{2}=2R_0$,
$\epsilon=0.1$, $\delta R=5$, $h=0.05$ and
$Q_\mathrm{out}=2$. An example of the initial surface density and the
Toomre $Q$ parameter is shown in
Fig. \ref{initial_surf}. Since $Q>1$, the disc is stable to local
axisymmetric perturbations \citep{toomre64}. 
The transition between self-gravitating and 
non-self-gravitating portions of the disc occur smoothly across
$\sim10H$. Initially there is no vertical or radial velocity
($v_R = v_r = v_\theta = 0$). The azimuthal velocity is initialized to
satisfy centrifugal balance with pressure and gravity,
\begin{align}
  \frac{v_\phi^2}{r} = \frac{1}{\rho}\frac{\p p}{\p r} + \frac{\p
    \Phi_\mathrm{tot}}{\p r}
\end{align}
and similarly in 2D.  

\subsection{Codes}
We use three independent grid-based codes to simulate the above 
system. We adopt computational units such that
$G=M_*=R_0=1$. Time is measured  in the Keplerian orbital period at
the reference radius, $P_0\equiv  2\pi/\Omega_k(R_0)$.   

\subsubsection{FARGO}
Our primary code is FARGO with self-gravity \citep{baruteau08}. This
is a popular, simple finite-difference code for 2D discs. `FARGO' refers
to its azimuthal transport algorithm, which removes the mean azimuthal
velocity of the disc, thereby permit larger time-steps than that would otherwise be allowed
by the usual Courant condition based on the full azimuthal
velocity \citep{masset00a,masset00b}.     

The 2D disc occupies
$R\in[R_\mathrm{min},R_\mathrm{max}],\,\phi\in[0,2\pi]$ and is  
divided into $(N_R,N_\phi)$ grids, logarithmically spaced in radius and
uniformly spaced in azimuth. At radial boundaries we set the
hydrodynamic variables to their initial values.   

The 2D Poisson equation is solved in integral form, 
\begin{align}\label{2d_grav}
  &\Phi_{d,z=0}(R,\phi) \notag \\
  &=-\int_{R_\mathrm{min}}^{R_\mathrm{max}} \int_0^{2\pi}
  \frac{G\Sigma(R^\prime,\phi^\prime)R^\prime dR^\prime d\phi^\prime}{\sqrt{R^2+R^{\prime 2} -
      2RR^\prime\cos{(\phi - \phi^\prime)} + \epsilon_g^2}}, 
\end{align}
using Fast Fourier Transform (FFT), where $\epsilon_g$ is a softening
length to prevent a numerical singularity. The FFT approach requires
$\epsilon_g\propto R$ \citep{baruteau08}. In FARGO, $\epsilon_g$ is
set to a fraction of $hR$.  

\subsubsection{ZEUS-MP}
ZEUS-MP  is a general-purpose finite difference
code \citep{hayes06}. We use the code in 3D spherical geometry, covering
$r\in[r_\mathrm{min},r_\mathrm{max}]$, $\theta\in[\theta_\mathrm{min},\pi/2]$,
$\phi\in[0,2\pi]$. The vertical domain is chosen to cover $n_H$
scale-heights at $R=R_0$, i.e. $\tan{(\pi/2 - \theta_\mathrm{min})}/h=n_H$. 
The grid is logarithmically spaced in radius and uniformly spaced in the angular
coordinates. We assume symmetry across the midplane, and
apply reflective boundary conditions at radial boundaries and the
upper disc boundary.  

ZEUS-MP solves the 3D Poisson equation using a conjugate gradient
method. To supply boundary conditions to the linear solver, we
expand the boundary potential in spherical harmonics $Y_{lm}$ 
as described in \cite{boss80}. The expansion is truncated at
$(l,m)=(\lmax,\mmax)$. This code was used in \cite{lin12b} for
self-gravitating disc-planet simulations.  

\subsubsection{PLUTO} 
PLUTO is a general-purpose Godunov code \citep{mignone07}. The grid
setup is the same as that adopted in ZEUS-MP above. We configure the
code similarly to that used in \cite{lin14}: piece-wise linear
reconstruction, a Roe solver and second order Runge-Kutta time
integration. We also enable the FARGO algorithm for azimuthal
transport. 

We solve the 3D Poisson equation throughout the domain using spherical
harmonic expansion \citep{boss80}, as used for the boundary potential
in ZEUS-MP. This version of PLUTO was used in \cite{lin14b} for
self-gravitating disc-planet simulations, producing similar results to
that of ZEUS-MP and FARGO.   

\subsection{Diagnostics}

\subsubsection{Evolution of non-axisymmetric modes}
The disc evolution is quantified using mode amplitudes and angular
momenta as follows. We list the 2D definitions with obvious 3D generalisations. 
A hydrodynamic variable $f$ (e.g. $\Sigma$) is written as 
\begin{align}
  f(R,\phi,t) &= \sum_{m=-\infty}^{\infty}f_m(R,t)\exp{\ii m \phi} \notag\\
  &= f_0 + 2 \real\left[\sum_{m=1}^\infty f_m \exp{(\ii
      m\phi)}\right], 
\end{align}
where the $f_m$ may be obtained from Fourier transform in $\phi$. 

The normalised surface density with azimuthal wavenumber $m$ is
\begin{align}
  \Delta\Sigma_m = \frac{2}{\Sigma_{00}} \real\left[\Sigma_m \exp{(\ii
      m\phi)}\right]
\end{align}
where $\Sigma_{00} = \Sigma_0(t=0)$. The time evolution of the
$m^\mathrm{th}$ mode can be characterized by 
assuming $\Sigma_m\propto\exp{(-\ii \sigma t)}$ as in linear
theory. The total non-axisymmetric surface density is 
\begin{align}
  \Delta\Sigma = \frac{\Sigma - \Sigma_0}{\Sigma_0}. 
\end{align}

\subsubsection{Angular momentum decomposition}
The total disc angular momentum is
\begin{align}
  J &= \int_{R_\mathrm{min}}^{R_\mathrm{max}}\int_0^{2\pi} \Sigma Rv_\phi RdRd\phi \notag\\
  &= 2\pi\int_{R_\mathrm{min}}^{R_\mathrm{max}} R\Sigma_0 v_{\phi0} R dR \notag\\ 
  &\phantom{=}+
  \sum_{m=1}^\infty2\pi\int_{R_\mathrm{min}}^{R_\mathrm{max}} 2R\real\left[\Sigma_m v_{\phi
      m}^*\right] RdR 
  = \sum_{m=0}^\infty J_m. 
\end{align}
We will refer to $J_m$ as the
$m^\mathrm{th}$ component of the total angular momentum, and use it to
monitor numerical angular momentum conservation in the simulations. 
It is important to distinguish this empirical definition from the
angular momentum of linear perturbations given in \S\ref{wkb}, which
is defined through a conservation law. 

\subsubsection{Three-dimensionality}
In 3D simulations we measure the importance of vertical motion with
$\Theta$, where 
\begin{align}\label{theta}
  \Theta^2 \equiv \frac{\avg{v_z^2}}{\avg{v_R^2}+\avg{v_\phi^2}}, 
\end{align}
and $\avg{\cdot}$ denotes the density-weighted average, e.g., 
\begin{align}
  \avg{v_z^2} \equiv\frac{
    \int_{R_1}^{R_2}\int_{\theta_\mathrm{min}}^{\pi/2} \int_{0}^{2\pi}
    \rho v_z^2 dV}{
    \int_{R_1}^{R_2}\int_{\theta_\mathrm{min}}^{\pi/2} \int_{0}^{2\pi}
    \rho dV
  },
\end{align}
and similarly for the horizontal velocities. Thus $\Theta$ is the
ratio of the average kinetic energy associated with vertical motion to
that in horizontal motion. The radial range of
integration is taken over $r\in[R_1,R_2]$ since this is where
we find the perturbations to be confined.

\section{Results}\label{results2d}
We first present results from FARGO simulations. The 2D disc spans
$[R_\mathrm{min}, R_\mathrm{max}] = [0.4,10]R_0$. This gives a total
disc mass $M_{d}=0.086M_*$. The mass within
$R\in[R_\mathrm{min},R_{1}]$ is $0.017M_*$, that within
$R\in[R_{1},R_{2}]$ is $0.049M_*$, and that within
$R\in[R_{2},R_\mathrm{max}]$ is $0.021M_*$. We use a resolution of
$N_R\times N_\phi = 1024\times 2048$, or about $16$ grids per $H$, and
adopt $\epsilon_g=10^{-4}hR$ for the   
self-gravity softening length\footnote{In 2D self-gravity, $\epsilon_g$ also
  approximates for the vertical disc thickness, so a more appropriate
  value would be $\epsilon_g\sim H$ \citep{muller12}. However, because
  $\epsilon_g\propto R$ is needed in FARGO, the Poisson kernel
  (Eq. \ref{2d_grav}) is no longer symmetric in $(R,R^\prime)$. We
  choose a small  
  $\epsilon_g$ in favour of angular momentum conservation, keeping in
  mind that the strength of self-gravity will be over-estimated.}.

In these simulations the disc is subject to initial perturbations in 
cylindrical radial velocity, 
\begin{align}\label{randpert}
  v_R \to v_R+ c_s\frac{\delta}{M}
  \exp{\left[-\frac{1}{2}\left(\frac{R-\overline{R}}{\Delta 
          R}\right)^2\right]}\sum_{m=1}^M\cos{m\phi},
\end{align}
where the amplitude $\delta\in[-10^{-3},10^{-3}]$ is set randomly but
independent of $\phi$, $\overline{R} = (R_{1}+R_{2})/2$
and $\Delta R = (R_{2}-R_{1})/2$. 

\subsection{Reference run}
\begin{figure}
  \includegraphics[width=\linewidth]{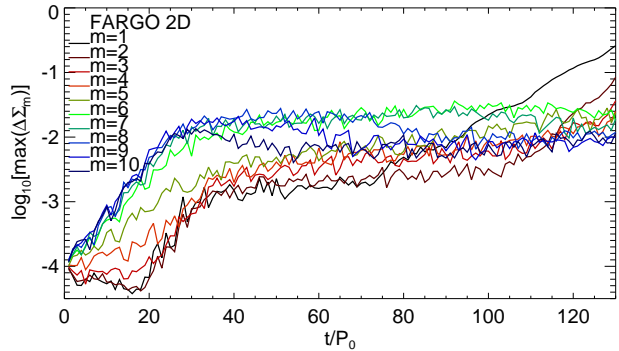}
  \caption{Evolution of non-axisymmetric surface density maxima 
    in the FARGO simulation initialised with perturbations
    with $m\in[1,10]$.\label{fargo_modeamp}} 
\end{figure}

\begin{figure*}
  \includegraphics[scale=0.55]{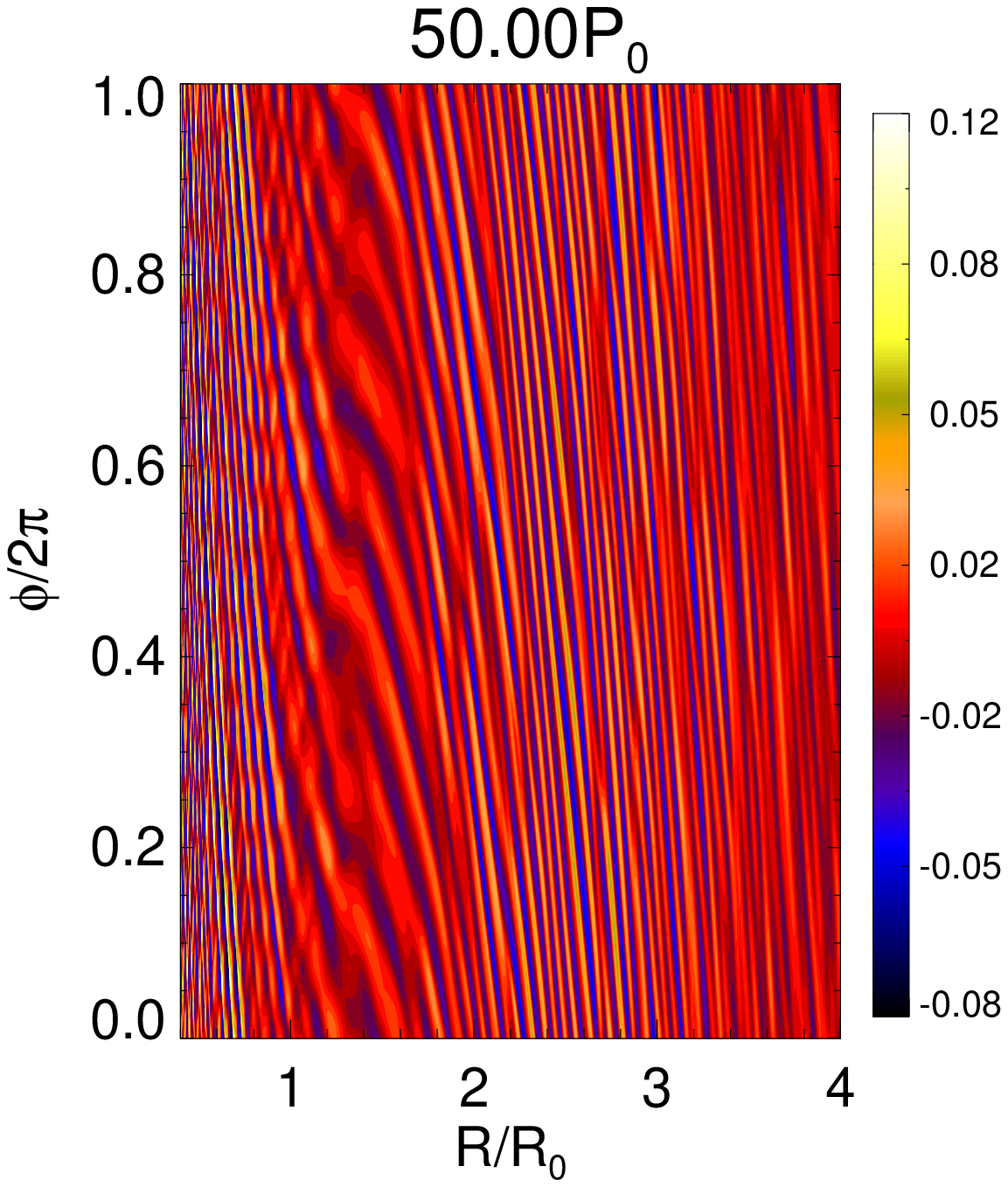}\includegraphics[scale=0.55,clip=true,trim=2.26cm 
  0cm 0cm 
  0cm]{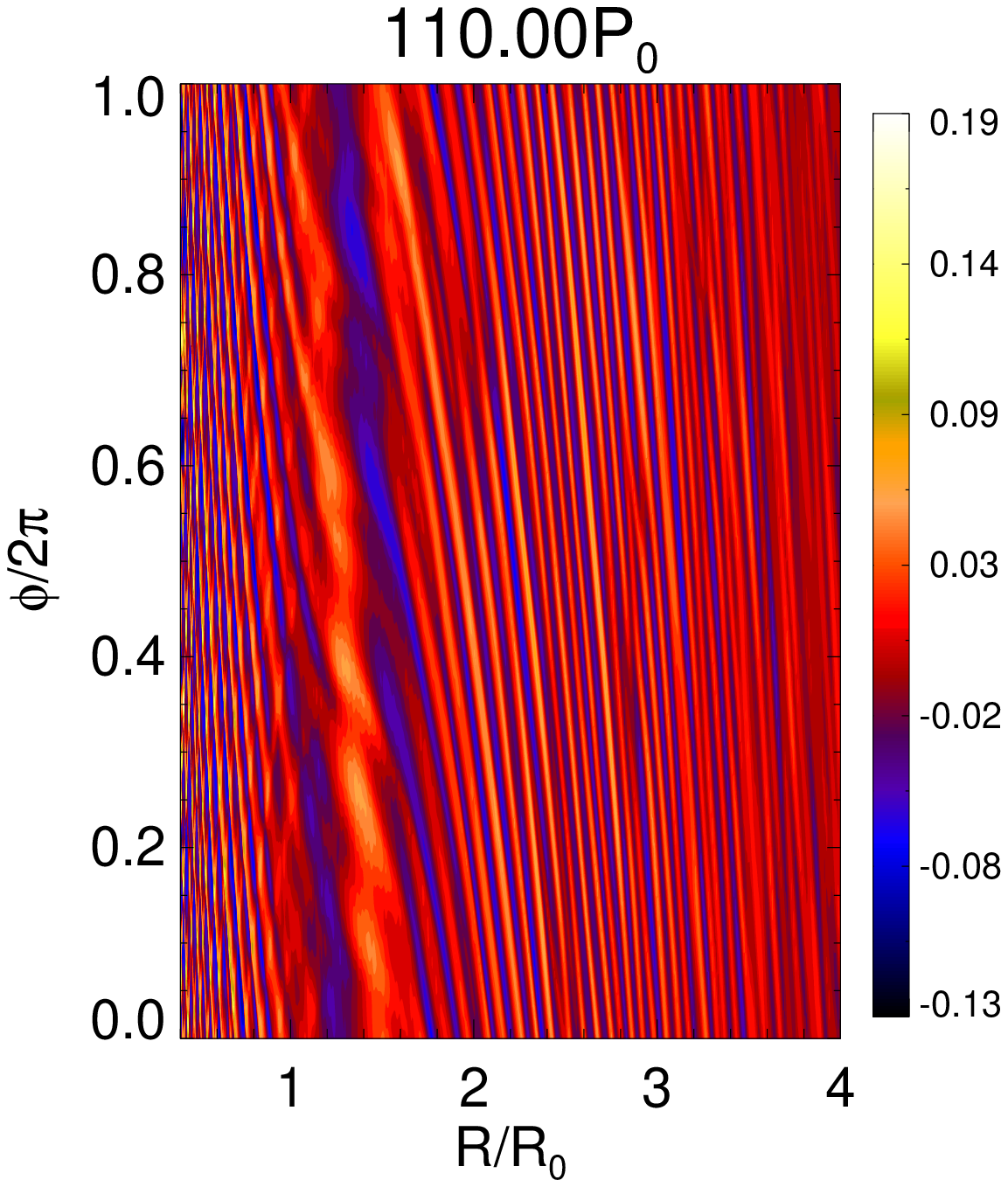}\includegraphics[scale=0.55,clip=true,trim=2.26cm
  0cm 0cm 0cm]{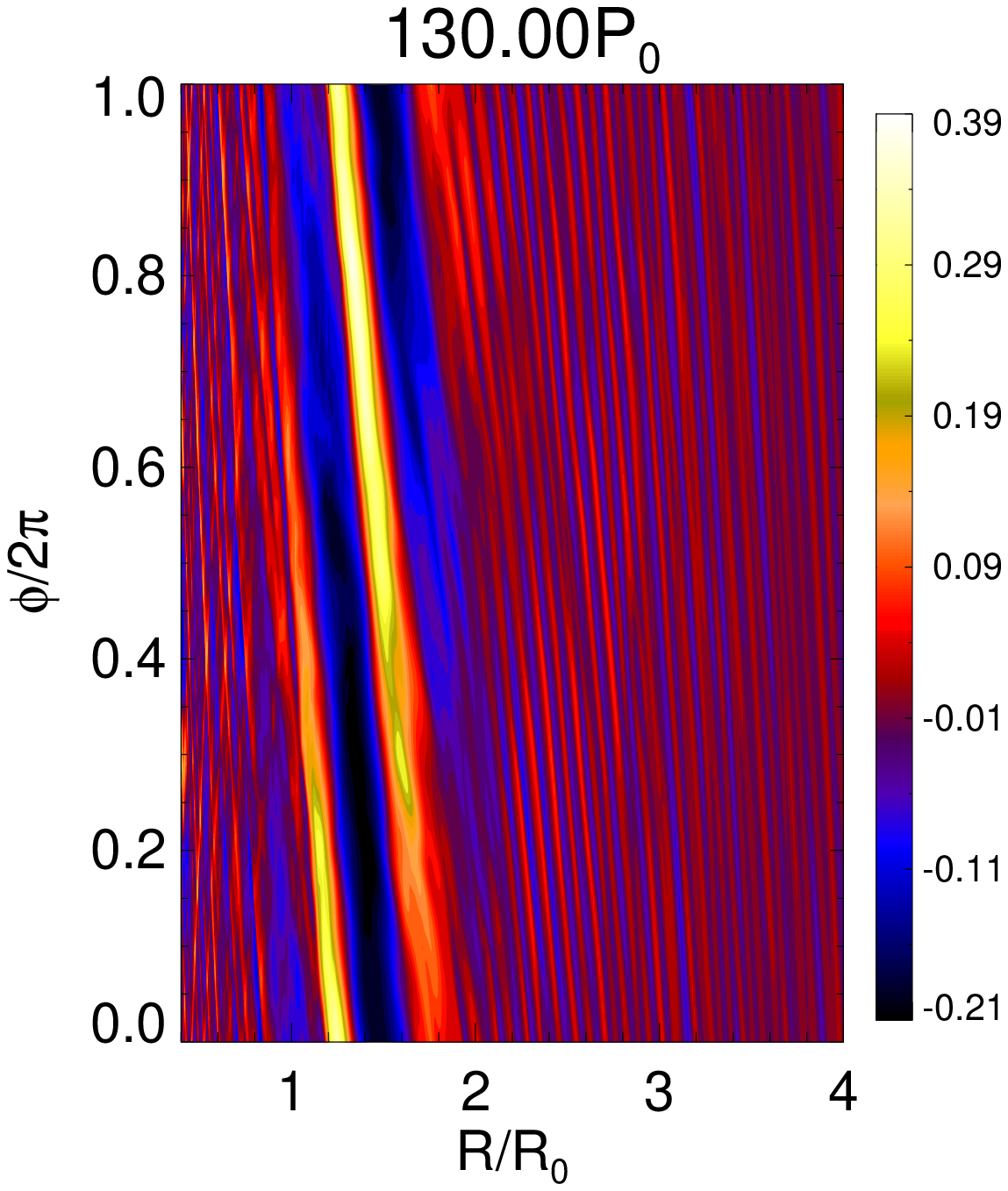} 
  \caption{Visualisation of the FARGO 2D simulation in
    Fig. \ref{fargo_modeamp}. The total  
    non-axisymmetric surface density
    $\Delta\Sigma$ is shown. \label{fargo_2d}} 
\end{figure*}

To obtain a picture of the overall disc evolution, we describe a 
fiducial run initialised with $M=10$ in Eq. \ref{randpert}. 
Fig. \ref{fargo_modeamp} plots evolution of the maximum 
non-axisymmetric surface density amplitudes in $R\in[R_{1},R_{2}]$
for $m\in[1,10]$. Snapshots from the simulation are shown in
Fig. \ref{fargo_2d}. 
At early times $t\lesssim100P_0$ the disc is  
dominated by low-amplitude high-$m$ perturbations. The $m\geq4$ modes
growth initially and saturate (or decays) after $t=40P_0$. Notice the
low $m\leq 2$ modes decay initially, but grows between $t\in[20,40]P_0$,
possibly due to non-linear interaction of the high-$m$ modes  
\citep{laughlin96,laughlin97}. However, the $m=1$ mode begins to grow
again after $t=70P_0$, and eventually dominates the annulus. 

\begin{figure}
  \includegraphics[width=\linewidth]{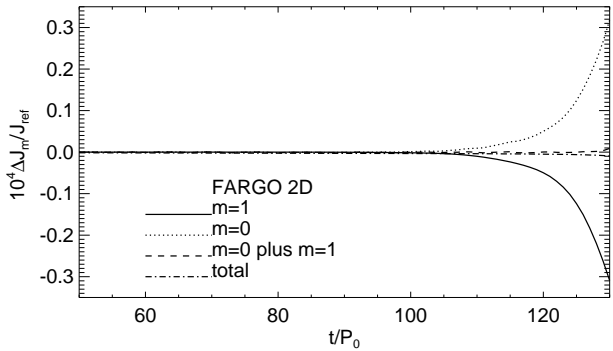}
  \caption{Evolution of angular momentum components in the 
    FARGO simulation in Fig. \ref{fargo_modeamp}---\ref{fargo_2d}. The
    perturbation relative to $t=0$ in 2D is shown in units of the 
    initial total angular momentum $J_\mathrm{ref}$.\label{2d_angmom}} 
\end{figure}

Fig. \ref{2d_angmom} shows the evolution of disc angular momentum
components. Only the $m=0,\,1$ components are 
plotted since they are dominant. The $m=1$ structure has
an associated negative angular momentum,  which indicates it is a
  low-frequency mode. 
Its growth is compensated by an increase in the axisymmetric
component of angular momentum, such that $\Delta J_0 + \Delta
J_1 \sim 0$. Note that FARGO does not conserve angular momentum
exactly. However, we find the total angular momentum varies by 
$|\Delta J/J|= O(10^{-6})$, and is much smaller than the
change in the angular momenta components, $|\Delta J_{0,1}/J|>
O(10^{-5})$. Fig. \ref{2d_angmom} then suggest that angular momentum
is transferred from the one-armed spiral to the background disc.    

\subsection{Dependence on the imposed temperature profile}
 We show that the growth of the $m=1$ spiral is
  associated with the imposed temperature gradient  by performing a 
  series of simulations with $q\in[0,1]$.  However, to maintain similar Toomre $Q$
  profiles, we adjust the surface density power-law index such
  that $s = (3+q)/2$.  For clarity these simulations are initialised
  with $m=1$ perturbations only.

\begin{figure}
  \includegraphics[width=\linewidth]{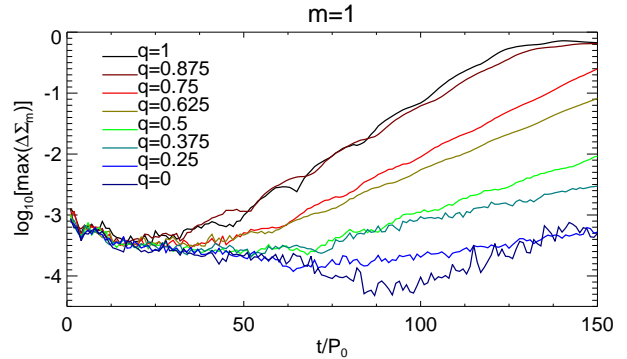}   
  \caption{Evolution of the $m=1$ spiral amplitude as a function of
    the negative of the imposed temperature gradient $q$. The maximum value of the
    $m=1$ surface density in $R\in[R_1,R_2]$ is shown. 
    \label{fargo_varq}} 
\end{figure}

Fig. \ref{fargo_varq} compares the $m=1$ spiral amplitudes as a
function of $q$. We indeed observe slower growth with decreasing
$q$. Although the figure indicates growth for the strictly isothermal 
disc ($q=0$), we did not observe a coherent one-armed spiral upon
inspection of the $m=1$ surface density field. The growth in this case
may be associated with high-$m$ modes, which dominated the 
simulation.

\begin{figure}
  \includegraphics[width=\linewidth]{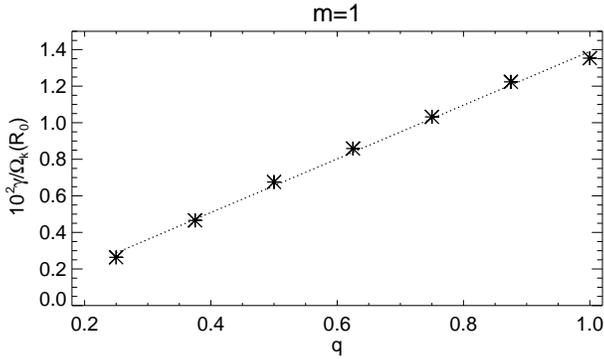}    
  \caption{Growth rates of the $m=1$ spiral mode as a function of the
    imposed sound-speed gradient $q$ (asterisks). A linear fit is also 
    plotted (dotted line). 
    \label{fargo_varq_growth}} 
\end{figure}

We plot growth rates of the $m=1$ mode as a function of $q$ in 
Fig. \ref{fargo_varq_growth}. The correlation can be fitted with a 
linear relation
\begin{align*}
  \gamma \simeq \left[0.015 q - 7.9\times10^{-4}\right] \Omega_k(R_0). 
\end{align*}
 As the background torque is proportional to $q$
  (Eq. \ref{theoretical_rate}), 
this indicates that the temperature gradient is responsible for the
development of the one-armed spirals observed in our simulations.  

We also performed a series of simulations with variable aspect-ratio 
$h\in[0.03,0.07]$ but fixed $q=1$. This affects the magnitude of the temperature
gradient since $c_s \propto h$. However, with other parameters equal
to that in the fiducial simulation,  varying $h$ also changes the disc
mass. For $h\in[0.03,0.07]$ the total disc mass ranges from
$M_d=0.052M_*$ to $M_d=0.12M_*$ and the 
mass within $R\in[R_1,R_2]$  ranges from $0.033M_*$ to $0.062M_*$. 

\begin{figure}
  \includegraphics[width=\linewidth]{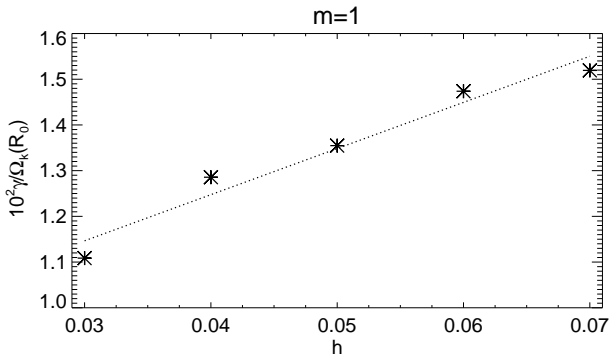}    
  \caption{Growth rates of the $m=1$ spiral mode as a function of the
    disc aspect-ratio $h$ (asterisks). A linear fit is also
    plotted (dotted line). 
    \label{fargo_varh_growth}} 
\end{figure}

Fig. \ref{fargo_varh_growth} shows the growth rates of the $m=1$
spiral in $R\in[R_1,R_2]$ as a function of $h$. Growth rates increases
with $h$, roughly as  
\begin{align*}
  \gamma \simeq \left[0.10h + 8.3\times10^{-3}\right]\Omega_k(R_0).  
\end{align*}
However, a linear fit is less good than for variable $q$ cases above. This
may be due to the change in the total disc mass when $h$ changes. We
find no qualitative difference between the spiral pattern that
emerges. 

\subsection{Properties of the $m=1$ spiral and its growth}\label{fargo_m1}
 Here we analyse the $q=1$ case in Fig. \ref{fargo_varq_growth} in
  more detail.  
 
\begin{figure}
  \includegraphics[width=\linewidth]{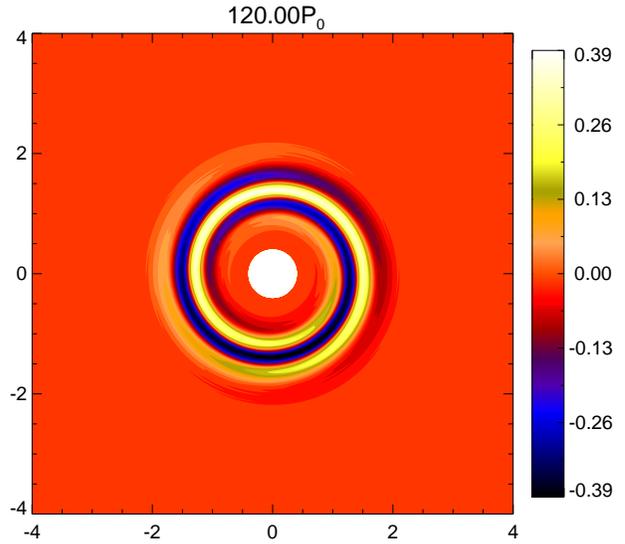}
  \caption{Cartesian visualisation of the $m=1$ surface density
    structure in the FARGO simulation initialised with only $m=1$
    perturbations. 
    \label{2d_fargo_viz}} 
\end{figure}

Fig. \ref{2d_fargo_viz} shows a snapshot of the $m=1$ surface
density of this run. 
By measuring the $m=1$ surface density amplitude and its
pattern speed, we obtain a co-rotation radius and growth rate 
\begin{align*}
  &R_c \simeq 4.4R_0,\\
  &\gamma\simeq 0.014\Omega_k(R_0) = 0.13\Omega_p. 
\end{align*}
This one-armed spiral can be considered as low frequency because its 
pattern speed $\Omega_p \simeq 0.1\Omega_k(R_0)\lesssim 0.3\Omega$ in
$R\in[R_{1},R_{2}]$ (where it has the largest amplitude). Thus, the spiral 
pattern appears nearly stationary. The growth rate $\gamma$ is also slow
relative to the local rotation, although the characteristic growth
time $\gamma^{-1} \simeq 10P_0$ is not very long.  

\begin{figure}
  \includegraphics[width=\linewidth]{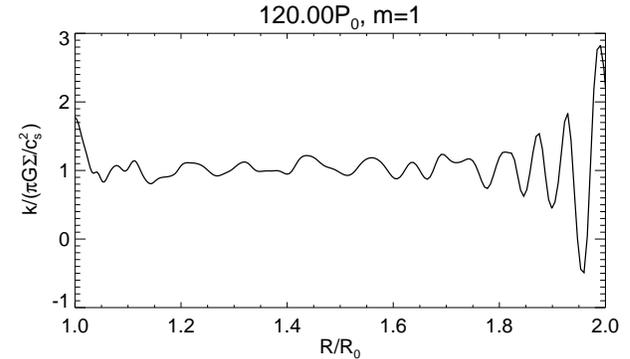}
  \caption{Normalised radial wavenumber of the $m=1$ spiral in 
    Fig. \ref{2d_fargo_viz}.\label{fargo_wavenumber}} 
\end{figure}

Next, we write $\Sigma_1 = 
|\Sigma_1|\exp{(\ii kR)}$, where $k$ is real, and assume the amplitude
$|\Sigma_1|$ varies slowly compared to the complex phase. This is the
main assumption in local theory. We calculate $k$ numerically and plot
its normalised value in Fig. \ref{fargo_wavenumber}. We find 
\begin{align*}
  kR \sim \frac{\pi G \Sigma}{c_s^2}R \sim \frac{1}{hQ}, 
\end{align*}
where we used $Q\sim c_s\Omega/\pi G \Sigma$ and $R\Omega/c_s\sim
h^{-1}$. Since $Q=O(1)$ and $h\ll 1$ imply $|kR|\gg 1$, we 
can apply results from local theory
  (\S\ref{local_approx}). Note also that $k\simeq k_c$. 
Fig. \ref{2d_fargo_viz} shows the $m=1$
spiral is trailing, consistent with $k>0$. 
 
\begin{figure}
  \includegraphics[width=\linewidth]{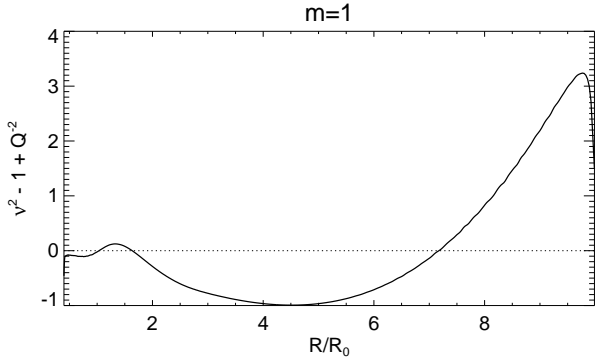} 
  \caption{Dimensionless mode frequency $\nu$ for the $m=1$ spiral in
    Fig. \ref{2d_fargo_viz}. For a given real mode frequency, the
    dispersion relation for local density waves, Eq. \ref{dispersion},
    permits purely wave-like solutions in regions where $\nu^2 - 1 +
    Q^{-2}>0$.    
    \label{fargo_qbarrier}} 
\end{figure}

Using the estimated value of $R_c$, we plot in
Fig. \ref{fargo_qbarrier} the quantity $\nu^2 - 1 + Q^{-2}$, which is
required to be positive in local theory for purely wave-like
solutions to the dispersion relation (Eq. \ref{dispersion}) when the
mode frequency is given. 
Fig. \ref{fargo_qbarrier} shows two $Q$-barriers located in
the inner disc, at $R_{Qb}=R_0$ and $R_{Qb}=1.6R_0$; the bounded region is indeed 
where the $m=1$ spiral develops. This shows that the one-armed 
spiral is trapped. Note in this region, $\nu^2 - 1 + 
Q^{-2}\simeq 0.1\ll 1$, which is necessary for consistency with 
the measured wavenumber $k$ and Eq. \ref{wavenumber}.  
There is one outer Lindblad resonance at $R_L\simeq
7.2R_0$. Thus, acoustic waves may be launched in $R\gtrsim 7.2R_0$ by
the spiral disturbance in the inner disc \citep{lin11b}. 

  We can estimate the expected growth rate of the $m=1$ mode due to the
  temperature gradient. Setting $k = k_c$ and $m=1$ into
  Eq. \ref{theoretical_rate} gives
  \begin{align}\label{theoretical_rate1}
    \gamma \sim \frac{qh}{2Q}\Omega. 
  \end{align}
  Inserting $q=1$, $h=0.05$ and $Q\simeq 1.5$  gives
  $\gamma \simeq 0.017\Omega$, consistent with numerical
  results.  

\subsubsection{Angular momentum exchange with the background disc}  
\begin{figure}
  \includegraphics[width=\linewidth]{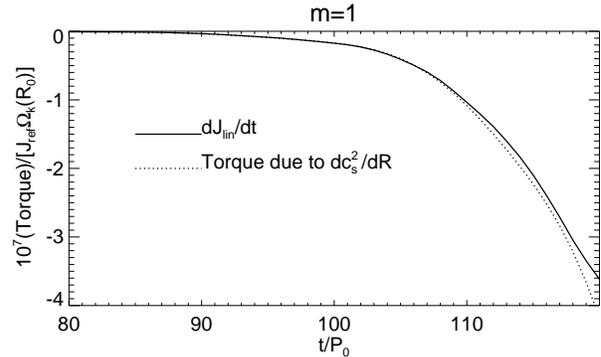} 
  \caption{Rate of change of the $m=1$ wave angular momentum as defined by
    Eq. \ref{baroclinic_torque_int} (solid) compared to the torque
    exerted on the wave associated with the background temperature
    gradient (dotted). 
    \label{fargo_angmom_ex}} 
\end{figure}

 We explicitly show that the growth of the $m=1$ spiral is due to
  the forced temperature gradient via the background torque described
  in \S\ref{wkb}. We integrate the statement for angular momentum 
  conservation for linear perturbations, Eq. \ref{lin_ang_mom_cons},
  assuming boundary fluxes are negligible, to obtain

\begin{align}\label{baroclinic_torque_int}
  \frac{d}{dt}\underbrace{\int_{R_\mathrm{min}}^{R_\mathrm{max}}\jlin
    2\pi R dR}_{\jlintot} 
  =\int_{R\mathrm{min}}^{R\mathrm{max}}T_\mathrm{BG} 2\pi R dR, 
\end{align}
where we recall $T_\mathrm{BG}$ is the torque density associated with the imposed
sound-speed profile (Eq. \ref{baroclinic_torque}). We  
compute both sides of Eq. \ref{baroclinic_torque_int} using
simulation data, and compare them in Fig. \ref{fargo_angmom_ex}. There
is a good match between the two torques, especially at early times
$t\lesssim110P_0$. The average discrepancy is $\simeq 5\%$. 
The match is less good later on, when the spiral
amplitude is no longer small ($\mathrm{max}\Delta\Sigma_1\sim 0.2$ at $t=110P_0$
and $\mathrm{max}\Delta\Sigma_1\sim 0.4$ by $t=120P_0$) and linear theory becomes
less applicable. Fig. \ref{fargo_angmom_ex} confirms that the $m=1$ spiral
wave experiences a negative torque that further reduces its (negative) angular
momentum, leading to its amplitude growth. This is consistent
with angular momentum component measurements (Fig. \ref {2d_angmom}).  

\section{Three-dimensional simulations}\label{results3d}
In this section we review 3D simulations carried 
out using ZEUS-MP and PLUTO. The main purpose is to verify 
the above results with different numerical codes, and validate  
the 2D approximation.    

The 3D disc has radial size
$[r_\mathrm{min},r_\mathrm{max}]=[0.4,10]R_0$ and vertical extent  
$n_H=2$ scale-heights at $R=R_0$. The resolution is $N_r\times N_\theta\times
N_\phi=256\times32\times512$, or about $4$ cells per
$H$. Because of the reduced resolution 
compared to 2D, we use a smooth perturbation by setting
$\delta = 10^{-3}$ and $M=1$ in Eq. \ref{randpert}. This corresponds
to a single $m=1$ disturbance in $R\in[R_1,R_2]$.

The 3D discs are initialised in approximate equilibrium only, so we
first evolve the disc without perturbations using  
$(\lmax,\mmax)=(32,0)$ up to $t=10P_0$, during which 
meridional velocities are damped out. We then restart the simulation
with the above perturbation and $(\lmax,\mmax)=(32,32)$, and damp
meridional velocities near the radial boundaries. 

\begin{figure}
  \includegraphics[width=\linewidth]{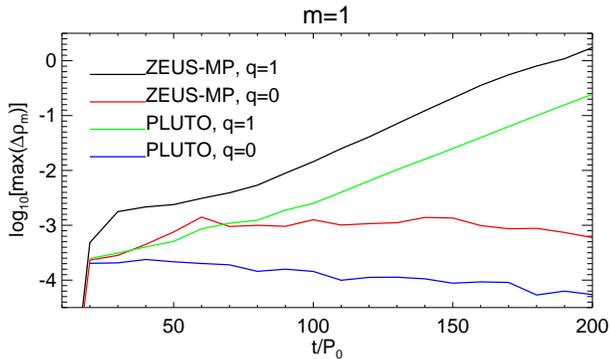}
  \caption{Evolution of the maximum $m=1$ density component in  $r\in[R_1,R_2]$
    in the 3D simulations. Results from discs with a 
    temperature gradient ($q=1$) and a strictly isothermal disc
    ($q=0$) are shown.  
    \label{3d_ampmax}}   
\end{figure}

Fig. \ref{3d_ampmax} plots the evolution of the $m=1$ spiral amplitudes measured
in the ZEUS-MP and PLUTO runs. We also ran simulations
with a strictly isothermal equation of state ($q=0$), which display no
growth compared to that with a temperature gradient.  This confirms
the temperature gradient effect is the same in 3D.

In the ZEUS-MP run, we observed high-$m$ disturbances developed near 
the inner boundary initially, which is likely responsible for the
growth seen at $t<50P_0$. This is a numerical artifact and effectively
seeds the simulation with a larger perturbation. Results
from ZEUS-MP are therefore off-set from PLUTO by $\sim50P_0$. However,
once the coherent $m=1$ spiral begins to grow ($t\gtrsim 100P_0$), 
we measure similar growth rates in both codes: 
\begin{align*}
  &\gamma \simeq 0.0073\Omega_k(R_0) \quad\quad \mathrm{PLUTO},\\
  &\gamma \simeq 0.0085\Omega_k(R_0) \quad\quad \varA{ZEUS-MP}.
\end{align*}
Both are somewhat smaller than the 2D simulations. This is
  possibly because of the lower resolutions adopted in 3D 
  and/or because the effective Toomre parameter is larger in 3D
  \citep{mamat10} which, from Eq. \ref{theoretical_rate1}, is
  stabilising. 


\begin{figure}
  \begin{center}
    \subfigure[ZEUS-MP]{
      \includegraphics[scale=0.305,clip=true,trim=0cm 0cm 0cm 0cm]{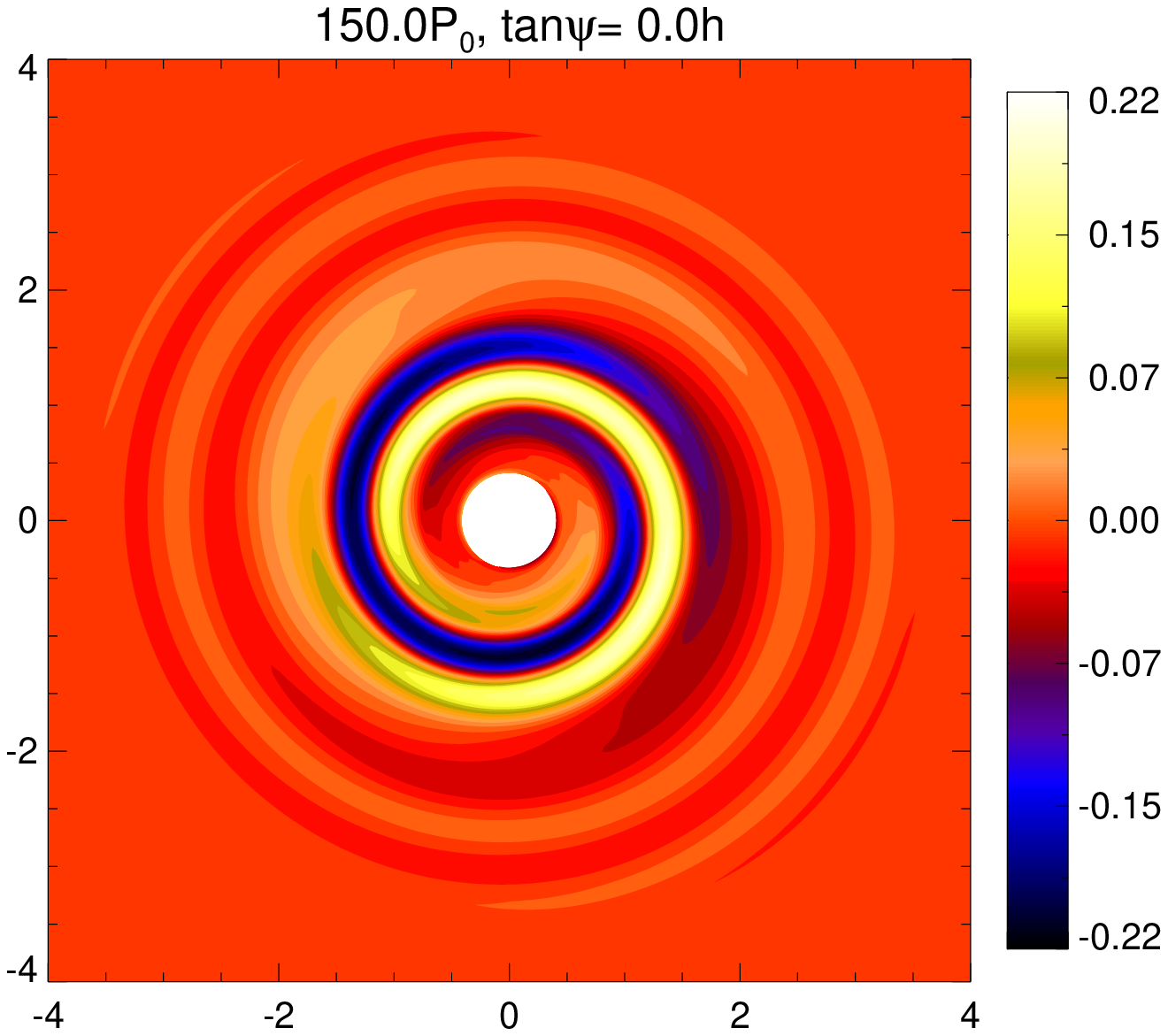}  
      \includegraphics[scale=0.305,clip=true,trim=0.8cm 0cm 0cm
      0cm]{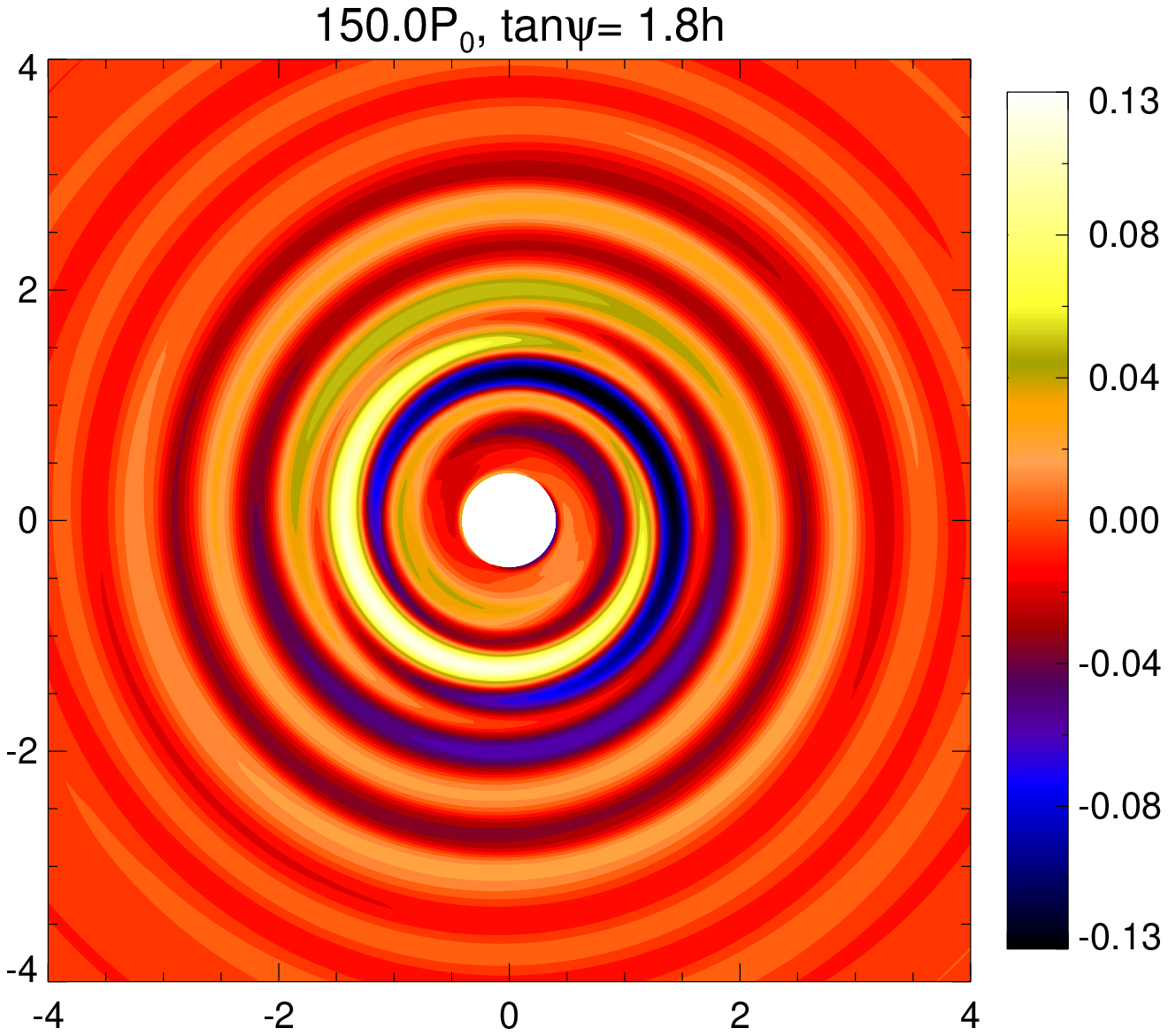}  
    }
    \subfigure[PLUTO]{
      \includegraphics[scale=0.305,clip=true,trim=0cm 0cm 0cm 0cm]{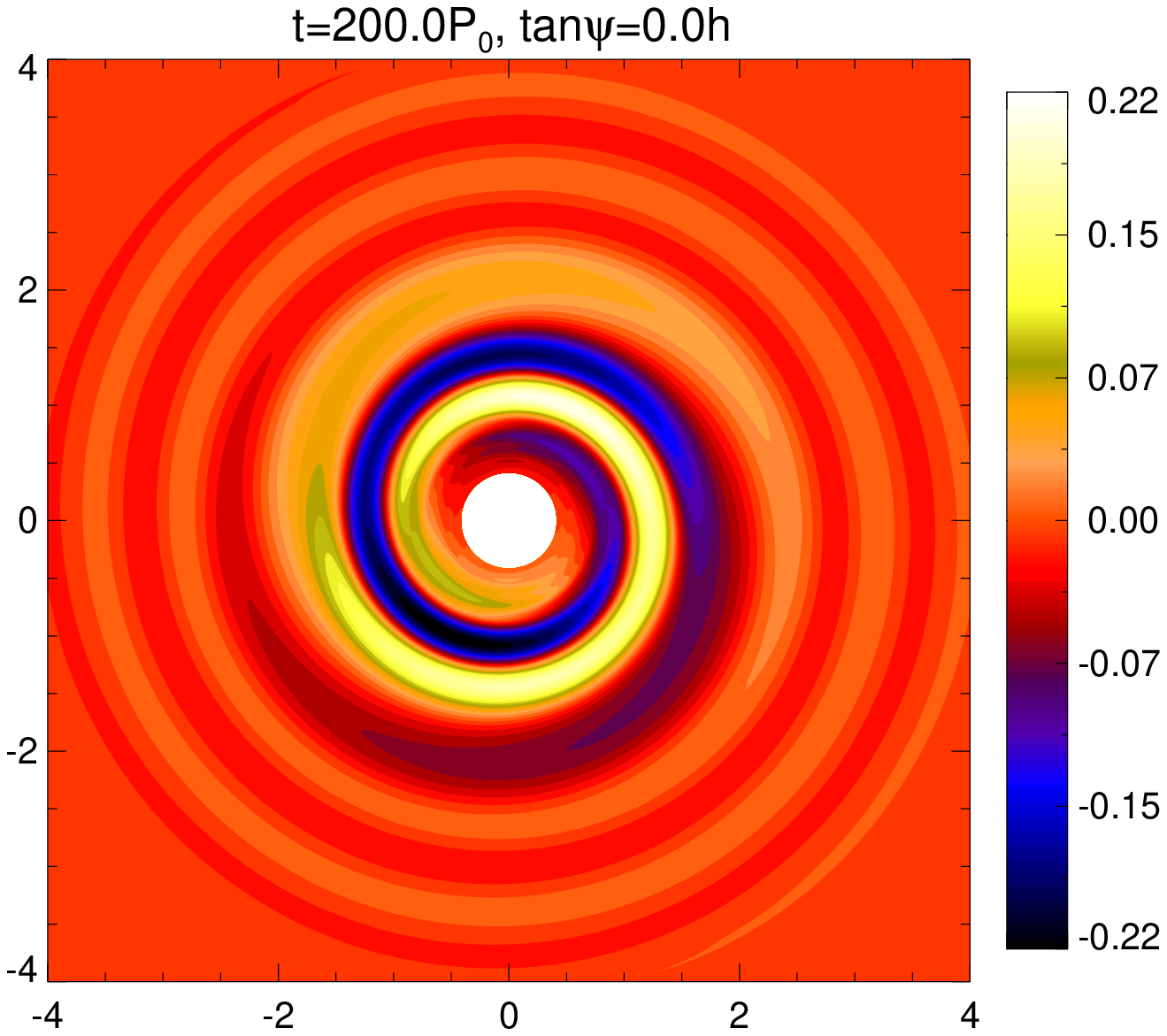}
      \includegraphics[scale=0.305,clip=true,trim=0.8cm 0cm 0cm
      0cm]{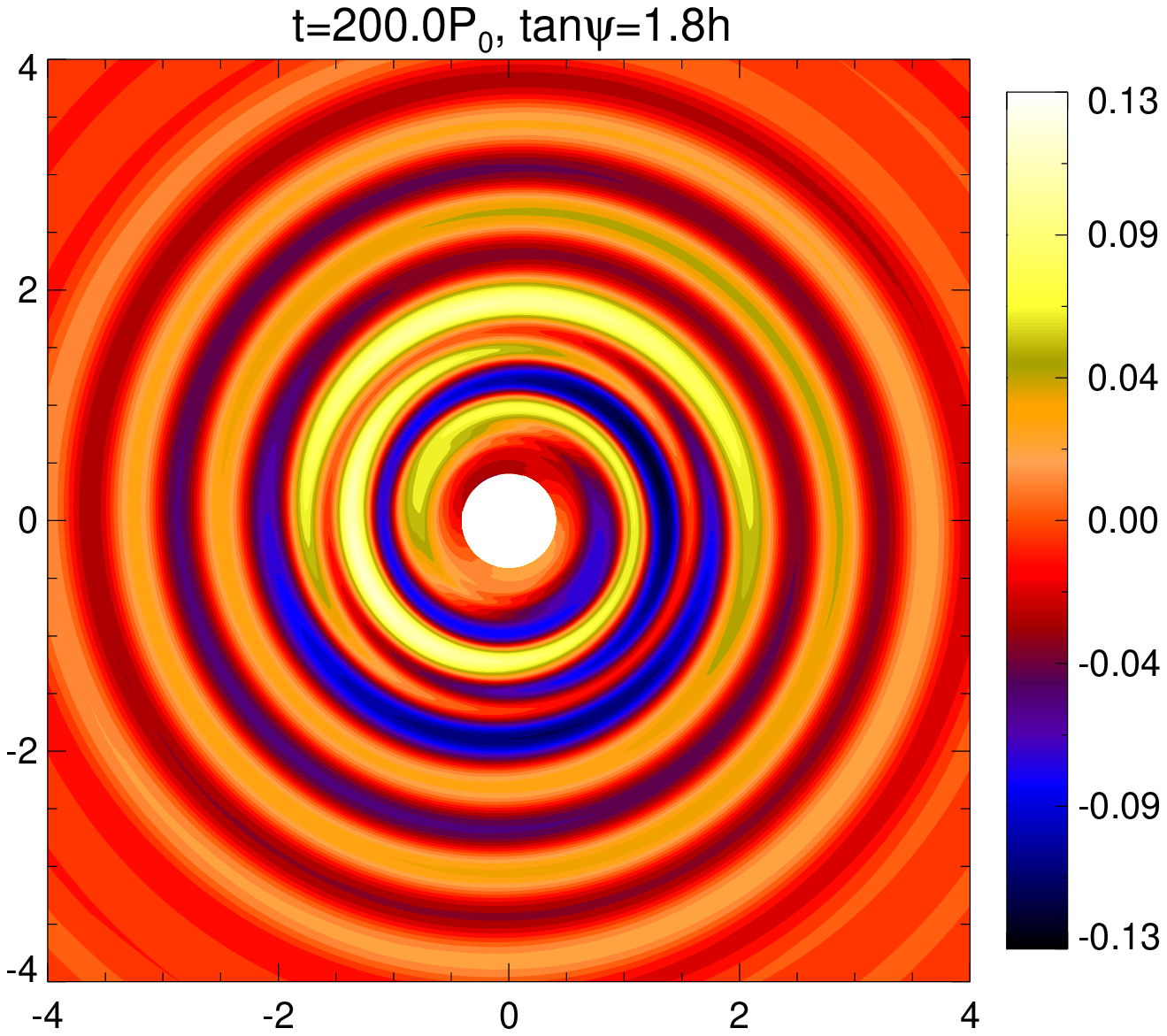}
    }
  \end{center}
  \caption{Three-dimensional simulations using the ZEUS-MP (top) and 
    PLUTO (bottom) codes. The $m=1$ density component $\Delta\rho_1$
    at the midplane (left) and approximately 
    two scale-heights above the midplane (right) is shown. Here $\psi
    \equiv \pi/2 - \theta$ is the angular height  
    from the midplane. \label{3d_prelim}}   
\end{figure}

Visualisations of the 3D simulations are shown in   
Fig. \ref{3d_prelim} for the disc midplane and near the upper disc
boundary. The snapshots are chosen when the one-armed spirals in the two codes
have reached comparable amplitudes. 
Both codes show similar one-armed patterns at either height, and the
midplane snapshot is similar to the 2D simulation
(Fig. \ref{fargo_2d}). 
The largest spiral amplitude is found in the
self-gravitating region $R\in[R_1,R_2]$, independent of
height. However, notice the spiral pattern extends into 
the non-self-gravitating outer disc ($R>R_2$) at $z\sim 2H$, i.e.    
the disturbance becomes more global away from the midplane.  

\subsection{Vertical structure}
\begin{figure}
  \includegraphics[scale=0.47,clip=true,trim=0cm 0.79cm 0cm
  0cm]{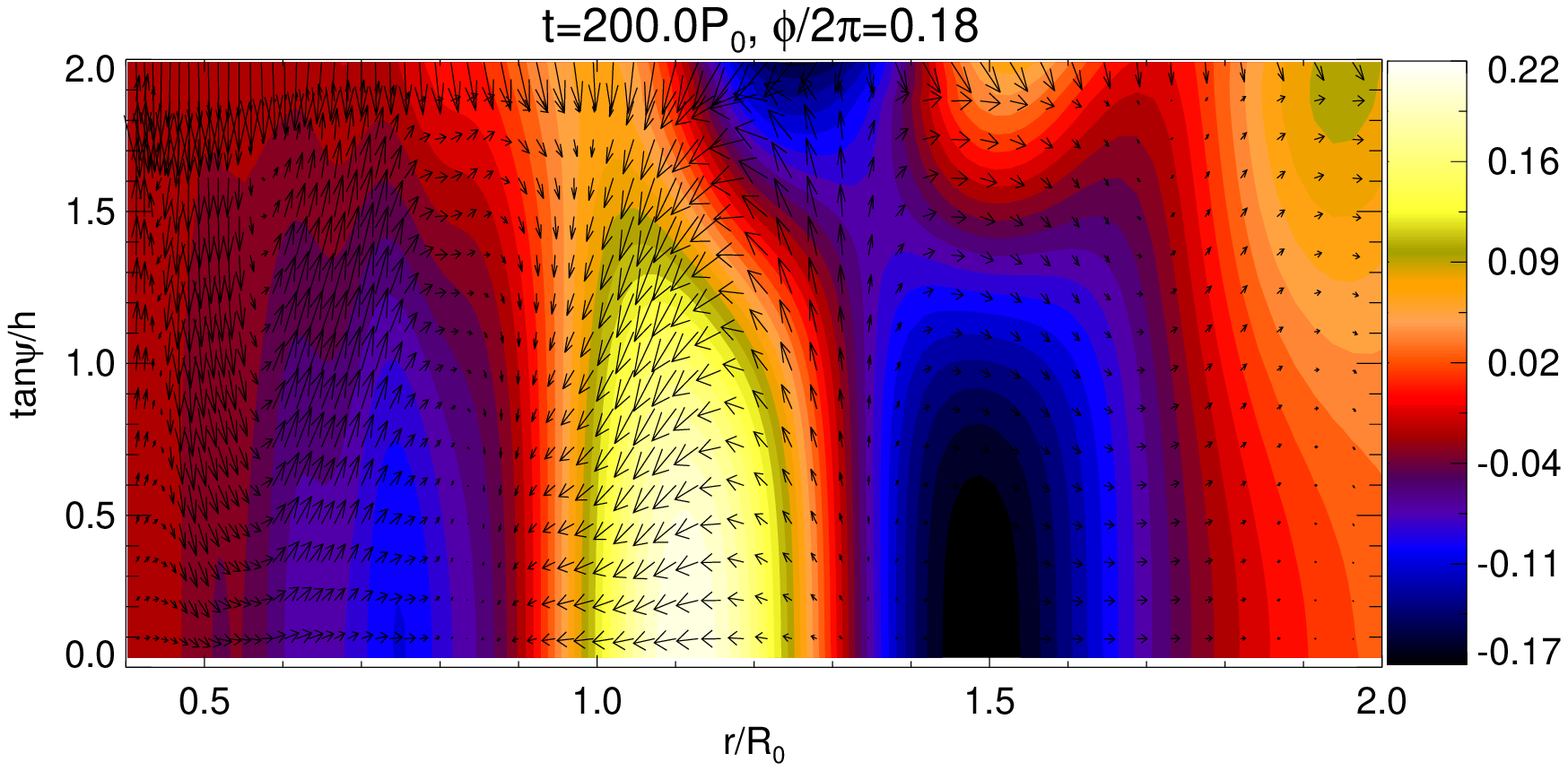}  
  \includegraphics[scale=0.47,clip=true,trim=0cm 0cm 0cm
  0.64cm]{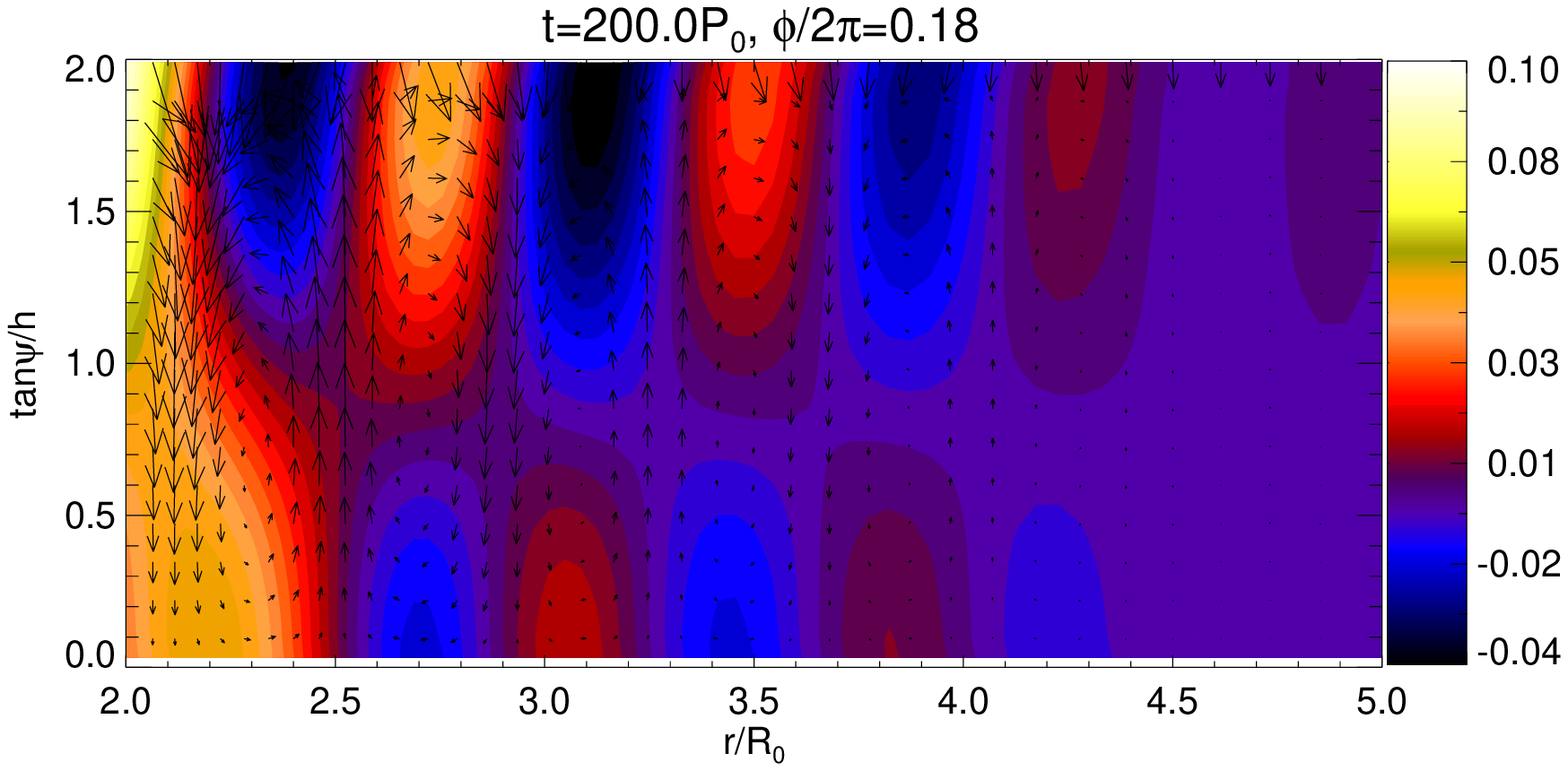}  
  \caption{The $m=1$ density component in the meridional plane in the 
    PLUTO simulation. The slices are taken at the azimuth of   
    $\mathrm{max}[\Delta\rho_1(r,\pi/2,\phi)]$. Arrows represent the vector 
    $(v_r/R_0,-v_\theta/rh\sin^2{\theta})$. The top (bottom) panel corresponds
    to the inner (outer) portions of the disc. 
    \label{3d_rz}} 
\end{figure}

Fig. \ref{3d_rz} shows the vertical structure of the one-armed 
spiral in the PLUTO run. The spiral is vertically confined to $z
\lesssim H$ at $R\sim R_0$ (the self-gravitating region). Thus, a 2D
disc model, representing dynamics near the disc midplane, is  
sufficient capture the instability. However, for $R>2R_0$ the spiral amplitude
increases away from the midplane. It remains  
small in our disc model ($|\Delta\rho_1| \lesssim 0.1$), but could become 
significant with a larger vertical domain. This means that 3D 
simulations are necessary to study the effect of the one-armed spiral
on the exterior disc.  
   
Although Fig. \ref{3d_rz} appears to display significant vertical motion,
we measure the three-dimensionality parameter $\Theta < 10^{-2}$
(Eq. \ref{theta}), so vertical motions are insignificant compared to
horizontal motions. This supports a 2D approximation. On the other
hand, we find  $\mathrm{max}|v_z/c_s|\sim 0.2$ which, although
sub-sonic, is not very small.  

\subsection{Angular momentum conservation}   
\begin{figure}
  \includegraphics[scale=.41,clip=true,trim=0cm 1cm 0cm 0cm]{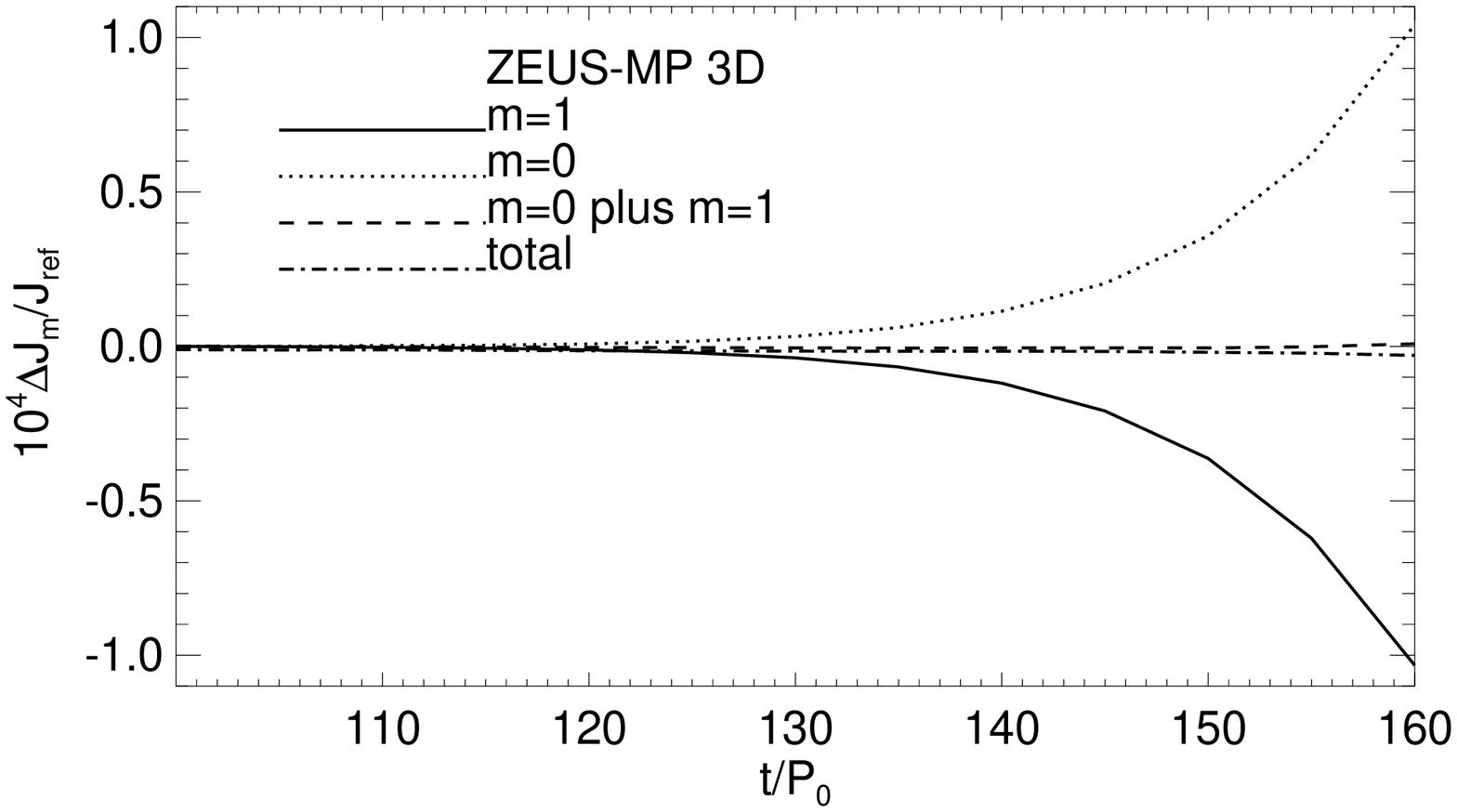}
  \includegraphics[scale=.41]{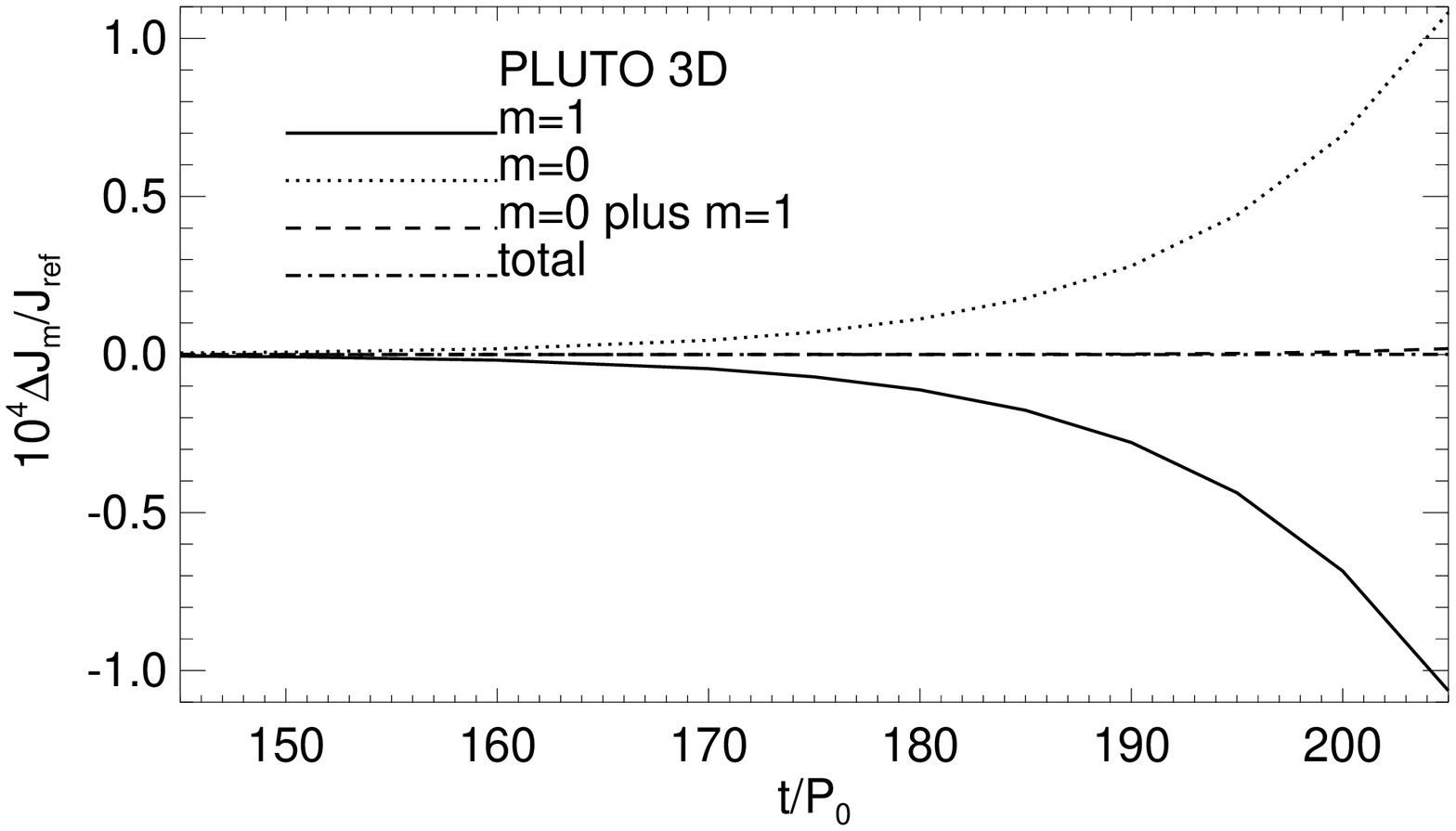}
  \caption{Evolution of angular momentum components in the 3D 
    simulations. The perturbation 
    relative to $t=10P_0$, during the growth of the one-armed spiral,
    is shown in units of the initial total angular momentum
    $J_\mathrm{ref}$.\label{3d_angmom}}  
\end{figure}

Fig. \ref{3d_angmom} shows the angular momentum evolution in the 3D 
runs during the linear growth of the one-armed spiral. Because the
ZEUS-MP simulation is off-set from PLUTO, the time interval for the
plot was chosen such that the change in the angular momentum
components are comparable in the two codes. 

ZEUS-MP does not conserve angular momentum very well, but the
variation in total angular momentum $|\Delta J/J|< O(10^{-6})$ is
small compared to the individual components $|\Delta J_{0,1}/J|\sim
10^{-4}$. 
PLUTO reaches similar values of $|\Delta J_{0,1}|$, but achieves better
conservation, with $|\Delta J/J|=O(10^{-8})$. These plots are again
similar to the 2D simulations, i.e. angular momentum lost by $J_1$ is
gained by $J_0$. This confirms that the interaction between $J_1$ and
$J_0$ operates in 3D and 2D similarly.  

\section{Discussion}\label{discussions} 

  We discuss below several issues related to self-gravitating
  discs in the context of our numerical simulations. However, it is
  important to keep in mind that 
  the growth of the one-armed spiral in our simulations is
  \emph{not} a gravitational instability in the sense that
  destabilisation is through the background torque associated with a
  forced temperature gradient, and not by self-gravitational torques\footnote{In fact, additional simulations with $Q_\mathrm{out}=4$ (giving $Q\gtrsim2.5$ throughout the disc) still 
  develops the one-armed spiral, but with a smaller growth rate.}.

  \subsection{Motion of the central star} 
  In our models we have purposefully neglected the indirect potential
  associated with a non-inertial reference frame to avoid
  complications arising from the motion of the central star.  
  Although it has been established that such motion can destabilise an
  $m=1$ disturbance in the disc \citep{adams89,shu90,michael10}, 
  the disc masses in our models ($M_d\lesssim 
  0.1 M_*$) are not expected to be sufficiently massive for this effect to be
  significant. Indeed, simulations including the
  indirect potential, carried out in the early stages of this project,
  produced similar results.    
  
  \subsection{Role of Lindblad and co-rotation torques}
  One effect of self-gravity is to allow the one-armed spiral,
  confined to $R\sim R_0$ in our models, to act as an external potential for the
  exterior disc in $R>R_0$. This is 
  analogous to disc-satellite interaction 
  \citep{goldreich79}, where the embedded satellite exerts a torque on
  the disc at Lindblad and co-rotation resonances. 

  In Appendix \ref{disc-planet} we estimate the magnitude of this effect
  using basic results from disc-planet theory \citep[see, e.g.][and
  references therein]{papaloizou07}. There, we find that the angular
  momentum exchange between the one-armed spiral and the exterior disc
  is insignificant compared to the background torque.  
  
  We confirmed this with additional FARGO simulations that exclude the
  co-rotation and outer Lindblad resonances (OLR) by reducing radial domain
  size to $R_\mathrm{max} = 3 R_0$, which still developed 
  the one-armed spiral. 

  \subsection{Applicability to protoplanetary discs}
  
  \subsubsection{Thermodynamic requirements}  

  A locally isothermal equation of state represents the ideal limit
  of infinitely short cooling and heating timescales, so the 
  disc temperature instantly returns to its initial value when 
  perturbed. The background torque is generally non-zero if the 
  resulting temperature profile has a non-zero radial gradient.  
  
  A short cooling timescale $t_c$ can occur in the outer
  parts of protoplanetary discs 
  \citep{rafikov05,clarke09,rice09,cossins10b,tsukamoto15}.  
  However, if a disc with $Q\simeq 1$ is cooled (towards zero
  temperature) on a timescale $t_c\lesssim\Omega_k^{-1}$, it will
  fragment following gravitational instability
  \citep{gammie01,rice05,paardekooper12}.  
  
  Fragmentation can be avoided if the disc is heated to maintain 
  $Q>Q_c$, the threshold for fragmentation \citep[$Q_c\simeq
  1.4$ for isothermal discs,][]{mayer04}. This may be 
  possible in the outer parts of protoplanetary discs due to
  stellar irradiation \citep{rafikov09,kratter11,zhu12}. Sufficiently strong
  external irradiation is expected to suppress the linear gravitational
  instabilities altogether \citep{rice11}.  

  The background torque may thus exist in the outer
  parts of protoplanetary discs that are irradiated, such
  that the disc temperature is set externally with a non-zero radial
  gradient \citep[e.g.][]{stamatellos08}. Of course, if external irradiation sets a strictly
  isothermal outer disc \citep[e.g.][]{boley09}, then the background
  torque vanishes.      


  
     
 
  \subsubsection{Radial disc structure}
  In our simulations the 
  $m=1$ spiral is confined between two $Q$-barriers, where real solutions to the local
  dispersion relation is possible (Eq. \ref{wavenumber}). The
  existence of such a cavity results from the adopted initial surface
  density bump (Eq. \ref{sig_bump}).  
  Thus, in our disc models the main role of disc structure 
  and self-gravity is to allow a local $m=1$ mode to be set up, which is then 
  destabilised by the background torque. 
  
  In order to confine an $m=1$ mode between two $Q$-barriers, we
  should have $Q^2(1-\nu^2)=1$ at two radii. Assuming 
  Keplerian rotation and a slow pattern speed $\Omega_p\ll\Omega$,
  this amounts to 
  \begin{align}\label{qb_cond}
    \left(\frac{R_{Qb}}{R_c}\right)^{-3/2} = 2Q^2(R_{Qb}). 
  \end{align}
  Then two $Q$-barriers may exist when the $Q^2$ profile
  rises more rapidly (decays more slowly) than $R^{-3/2}$
  for decreasing (increasing) $R$. Note that 
  Eq. \ref{qb_cond} does not necessarily require strong self-gravity
  if $R_c$ is large. 
  
  A surface density bump can develop in 
  `dead zones' of protoplanetary 
  discs, where there is reduced mass accretion because the magneto-rotational 
  instability is ineffective for angular momentum transport 
  \citep{gammie96,turner08,landry13}. The dead zone becomes 
  self-gravitating with sufficient mass built-up 
  \citep{armitage01,martin12,martin12b,zhu09,zhu10,zhu10b,bae13}.  
  
  However, conditions in a dead zone may not be suitable for 
  sustaining a background torque because it may not cool/heat fast enough
  to maintain a fixed temperature profile. Recently, 
  \cite{bae14} presented numerical models of
  dead zones including a range of heating and 
  cooling processes, which show that dead zones developed large-scale
  (genuine) gravitational instabilities with multiple spiral
  arms. Although this does not prove absence of the background torque,
  it is probably insignificant compared to gravitational torques. 

  Another possibility is a gap opened by an embedded planet. In that case $Q$
  rises rapidly towards the gap centre since it is a region of low
  surface density. This can satisfy Eq. \ref{qb_cond}. Then the inner
  edge of our bump function mimics the outer gap edge. The outer gap
  edge is then a potential site for the growth of a low-frequency
  one-armed spiral through the background torque. However, the
  locally isothermal requirement would limit this process to the
  outer disc, or that the temperature 
  profile about the gap edge is set by the planet luminosity.  

  Here, it is worth mentioning the transition disc 
  around HD 142527, the outer parts of which displays an $m=1$
  asymmetry \citep{fukagawa13} and spiral arms 
  \citep{christiaens14} just outside a disc gap. These authors estimate 
  $Q\simeq$1---2 in the outer disc, implying self-gravity is
  important, but the disc may remain gravitationally-stable
  \citep{christiaens14}. This is a  
  possible situation that our disc models represent.


\section{Summary and conclusions}\label{summary}

  In this paper, we have described a destabilising  
  effect of adopting a fixed temperature profile to model   
  astrophysical discs. By applying angular momentum conservation
  within linear theory, we showed that a forced temperature gradient 
  introduces a torque on linear perturbations. We call this the
  background torque because it represents an exchange of angular
  momentum between the background disc and the perturbations. This 
  offers a previously unexplored pathway to instability in locally
  isothermal discs.

In the local approximation, we showed that this background torque is
negative for non-axisymmetric trailing waves in discs with a fixed temperature or
sound-speed profile that decrease outwards. A negative background torque enforces  
low-frequency non-axisymmetric modes because they are associated
with negative angular momentum.

  We demonstrated the destabilising effect of the background torque by
  carrying out direct numerical hydrodynamic simulations of 
  locally isothermal discs with a self-gravitating surface density
  bump. 
We find such systems are unstable to low-frequency perturbations
with azimuthal wavenumber $m=1$, which leads to the development of an one-armed 
trailing spiral that persist for at least $O(10^2)$ orbits. The spiral
pattern speed is smaller than the local disc rotation and
growth rates are $O(10^{-2}\Omega)$ which gives a characteristic
growth time of $O(10)$ orbits. 

We used three independent numerical codes --- FARGO in 2D, ZEUS-MP and
PLUTO in 3D --- to show that the growth of     
one-armed spirals in our disc model is due to the imposed
temperature gradient: growth rates increased linearly   
with the magnitude of the imposed temperature gradient, and one-armed
spirals did not develop in strictly isothermal simulations. This 
  one-armed spiral instability can be interpreted as an initially 
  neutral, tightly-wound $m=1$ mode being destabilised by the 
  background torque. The spiral
is mostly confined between two $Q$-barriers in the surface density bump. 
We find the instability behaves similarly in 2D and 3D, but in 3D
the spiral disturbance becomes more radially global away from    
the midplane. 

\subsection{Speculations and future work}

 There are several issues that remain to be addressed in future
  works:  

 \emph{Thermal relaxation.} The locally isothermal assumption 
 can be relaxed by including an energy equation with
 a source term that restores the disc temperature over a
  characteristic timescale $t_\mathrm{relax}$. Preliminary FARGO
simulations indicate a thermal relaxation timescale $t_\mathrm{relax} <
0.1\Omega_k^{-1}$ is needed for the one-armed spiral to
develop. However, this value is likely model-dependent. For example,
a longer $t_\mathrm{relax}$ may be permitted with larger temperature
gradients. This issue, together with a parameter survey, will be
considered in a follow-up study.  


 \emph{Non-linear evolution}. In the deeply non-linear regime, the
one-armed spiral may  
shock and deposit negative angular momentum onto 
the background disc. The spiral amplitude would saturate by gaining
positive angular momentum. However, if the temperature gradient is
maintained, it may be possible to achieve a balance between the gain
of negative angular momentum through the background torque, and the
gain of positive angular momentum through shock dissipation. We remark  
that fragmentation is unlikely because the co-rotation radius is
outside the bulk of the spiral arm \citep{durisen08,rogers12}. In order
to study these possibilities, improved numerical models are needed to
ensure total angular momentum conservation on timescales much longer
than that considered in this paper. 



\emph{Other applications of the background torque.}     
 The background torque is a generic feature in  
  discs for which the temperature is set externally.  It may therefore
  be relevant in other astrophysical contexts. 
One possibility is in Be star discs \citep{rivinius13}, 
for which one-armed oscillations may explain long-timescale variations
in their emission lines \citep[see e.g.][and references 
therein]{okasaki97,papaloizou06c,ogilvie08}. These studies 
invoke alternative mechanisms to produce \emph{neutral} one-armed
oscillations (e.g. rotational deformation of the star), but consider 
strictly isothermal discs. It would be interesting to explore
the effect of a radial temperature gradient on the stability of these
oscillations.


\section*{Acknowledgments}
I thank K. Kratter, Y. Wu and A. Youdin for valuable discussions, and the
anonymous referee for comments that significantly improved this paper. All
computations were performed on the El Gato cluster at the University
of Arizona. This material is based upon work supported by the National
Science Foundation under Grant No. 1228509. 

\bibliographystyle{mn2e}
\bibliography{ref}

\begin{thebibliography}{93}
\expandafter\ifx\csname natexlab\endcsname\relax\def\natexlab#1{#1}\fi

\bibitem[{{Adams}, {Ruden} \& {Shu}(1989){Adams}, {Ruden}, \& {Shu}}]{adams89}
{Adams} F.~C., {Ruden} S.~P., {Shu} F.~H., 1989, \apj, 347, 959

\bibitem[{{Armitage}, {Livio} \& {Pringle}(2001){Armitage}, {Livio}, \&
  {Pringle}}]{armitage01}
{Armitage} P.~J., {Livio} M., {Pringle} J.~E., 2001, \mnras, 324, 705

\bibitem[{{Avenhaus} {et~al}\mbox{.}(2014){Avenhaus}, {Quanz}, {Schmid},
  {Meyer}, {Garufi}, {Wolf}, \& {Dominik}}]{avenhaus14}
{Avenhaus} H., {Quanz} S.~P., {Schmid} H.~M., {Meyer} M.~R., {Garufi} A.,
  {Wolf} S., {Dominik} C., 2014, \apj, 781, 87

\bibitem[{{Bae} {et~al}\mbox{.}(2013){Bae}, {Hartmann}, {Zhu}, \&
  {Gammie}}]{bae13}
{Bae} J., {Hartmann} L., {Zhu} Z., {Gammie} C., 2013, \apj, 764, 141

\bibitem[{{Bae} {et~al}\mbox{.}(2014){Bae}, {Hartmann}, {Zhu}, \&
  {Nelson}}]{bae14}
{Bae} J., {Hartmann} L., {Zhu} Z., {Nelson} R.~P., 2014, \apj, 795, 61

\bibitem[{{Balbus} \& {Papaloizou}(1999)}]{balbus99}
{Balbus} S.~A., {Papaloizou} J.~C.~B., 1999, \apj, 521, 650

\bibitem[{{Baruteau} {et~al}\mbox{.}(2013){Baruteau}, {Crida}, {Paardekooper},
  {Masset}, {Guilet}, {Bitsch}, {Nelson}, {Kley}, \&
  {Papaloizou}}]{baruteau13b}
{Baruteau} C. {et~al.}, 2013, ArXiv e-prints

\bibitem[{{Baruteau} \& {Masset}(2008)}]{baruteau08}
{Baruteau} C., {Masset} F., 2008, \apj, 678, 483

\bibitem[{{Boccaletti} {et~al}\mbox{.}(2013){Boccaletti}, {Pantin}, {Lagrange},
  {Augereau}, {Meheut}, \& {Quanz}}]{boccaletti14}
{Boccaletti} A., {Pantin} E., {Lagrange} A.-M., {Augereau} J.-C., {Meheut} H.,
  {Quanz} S.~P., 2013, \aap, 560, A20

\bibitem[{{Boley}(2009)}]{boley09}
{Boley} A.~C., 2009, \apjl, 695, L53

\bibitem[{{Boss}(1980)}]{boss80}
{Boss} A.~P., 1980, \apj, 236, 619

\bibitem[{{Casassus} {et~al}\mbox{.}(2013){Casassus}, {van der Plas}, {M},
  {Dent}, {Fomalont}, {Hagelberg}, {Hales}, {Jord{\'a}n}, {Mawet},
  {M{\'e}nard}, {Wootten}, {Wilner}, {Hughes}, {Schreiber}, {Girard},
  {Ercolano}, {Canovas}, {Rom{\'a}n}, \& {Salinas}}]{casassus13}
{Casassus} S. {et~al.}, 2013, \nat, 493, 191

\bibitem[{{Christiaens} {et~al}\mbox{.}(2014){Christiaens}, {Casassus},
  {Perez}, {van der Plas}, \& {M{\'e}nard}}]{christiaens14}
{Christiaens} V., {Casassus} S., {Perez} S., {van der Plas} G., {M{\'e}nard}
  F., 2014, \apjl, 785, L12

\bibitem[{{Clarke}(2009)}]{clarke09}
{Clarke} C.~J., 2009, \mnras, 396, 1066

\bibitem[{{Cossins}, {Lodato} \& {Clarke}(2010){Cossins}, {Lodato}, \&
  {Clarke}}]{cossins10b}
{Cossins} P., {Lodato} G., {Clarke} C., 2010, \mnras, 401, 2587

\bibitem[{{Cossins}, {Lodato} \& {Testi}(2010){Cossins}, {Lodato}, \&
  {Testi}}]{cossins10}
{Cossins} P., {Lodato} G., {Testi} L., 2010, \mnras, 407, 181

\bibitem[{{Dipierro} {et~al}\mbox{.}(2014){Dipierro}, {Lodato}, {Testi}, \& {de
  Gregorio Monsalvo}}]{dipierror14}
{Dipierro} G., {Lodato} G., {Testi} L., {de Gregorio Monsalvo} I., 2014,
  \mnras, 444, 1919

\bibitem[{{Durisen}, {Hartquist} \& {Pickett}(2008){Durisen}, {Hartquist}, \&
  {Pickett}}]{durisen08}
{Durisen} R.~H., {Hartquist} T.~W., {Pickett} M.~K., 2008, \apss, 317, 3

\bibitem[{{Follette} {et~al}\mbox{.}(2014){Follette}, {Grady}, {Swearingen},
  {Sitko}, {Champney}, {van der Marel}, {Takami}, {Kuchner}, {Close}, {Muto},
  {Mayama}, {McElwain}, {Fukagawa}, {Maaskant}, {Min}, {Russell}, {Kudo},
  {Kusakabe}, {Hashimoto}, {Abe}, {Akiyama}, {Brandner}, {Brandt}, {Carson},
  {Currie}, {Egner}, {Feldt}, {Goto}, {Guyon}, {Hayano}, {Hayashi}, {Hayashi},
  {Henning}, {Hodapp}, {Ishii}, {Iye}, {Janson}, {Kandori}, {Knapp},
  {Kuzuhara}, {Kwon}, {Matsuo}, {Miyama}, {Morino}, {Moro-Martin}, {Nishimura},
  {Pyo}, {Serabyn}, {Suenaga}, {Suto}, {Suzuki}, {Takahashi}, {Takato},
  {Terada}, {Thalmann}, {Tomono}, {Turner}, {Watanabe}, {Wisniewski}, {Yamada},
  {Takami}, {Usuda}, \& {Tamura}}]{follette14}
{Follette} K.~B. {et~al.}, 2014, ArXiv e-prints

\bibitem[{{Forgan} {et~al}\mbox{.}(2011){Forgan}, {Rice}, {Cossins}, \&
  {Lodato}}]{forgan11}
{Forgan} D., {Rice} K., {Cossins} P., {Lodato} G., 2011, \mnras, 410, 994

\bibitem[{{Fukagawa} {et~al}\mbox{.}(2013){Fukagawa}, {Tsukagoshi}, {Momose},
  {Saigo}, {Ohashi}, {Kitamura}, {Inutsuka}, {Muto}, {Nomura}, {Takeuchi},
  {Kobayashi}, {Hanawa}, {Akiyama}, {Honda}, {Fujiwara}, {Kataoka},
  {Takahashi}, \& {Shibai}}]{fukagawa13}
{Fukagawa} M. {et~al.}, 2013, \pasj, 65, L14

\bibitem[{{Gammie}(1996)}]{gammie96}
{Gammie} C.~F., 1996, \apj, 457, 355

\bibitem[{{Gammie}(2001)}]{gammie01}
{Gammie} C.~F., 2001, \apj, 553, 174

\bibitem[{{Goldreich} \& {Lynden-Bell}(1965)}]{goldreich65}
{Goldreich} P., {Lynden-Bell} D., 1965, \mnras, 130, 125

\bibitem[{{Goldreich} \& {Tremaine}(1979)}]{goldreich79}
{Goldreich} P., {Tremaine} S., 1979, \apj, 233, 857

\bibitem[{{Grady} {et~al}\mbox{.}(2013){Grady}, {Muto}, {Hashimoto},
  {Fukagawa}, {Currie}, {Biller}, {Thalmann}, {Sitko}, {Russell}, {Wisniewski},
  {Dong}, {Kwon}, {Sai}, {Hornbeck}, {Schneider}, {Hines}, {Moro
  Mart{\'{\i}}n}, {Feldt}, {Henning}, {Pott}, {Bonnefoy}, {Bouwman}, {Lacour},
  {Mueller}, {Juh{\'a}sz}, {Crida}, {Chauvin}, {Andrews}, {Wilner}, {Kraus},
  {Dahm}, {Robitaille}, {Jang-Condell}, {Abe}, {Akiyama}, {Brandner}, {Brandt},
  {Carson}, {Egner}, {Follette}, {Goto}, {Guyon}, {Hayano}, {Hayashi},
  {Hayashi}, {Hodapp}, {Ishii}, {Iye}, {Janson}, {Kandori}, {Knapp}, {Kudo},
  {Kusakabe}, {Kuzuhara}, {Mayama}, {McElwain}, {Matsuo}, {Miyama}, {Morino},
  {Nishimura}, {Pyo}, {Serabyn}, {Suto}, {Suzuki}, {Takami}, {Takato},
  {Terada}, {Tomono}, {Turner}, {Watanabe}, {Yamada}, {Takami}, {Usuda}, \&
  {Tamura}}]{grady13}
{Grady} C.~A. {et~al.}, 2013, \apj, 762, 48

\bibitem[{{Hashimoto} {et~al}\mbox{.}(2011){Hashimoto}, {Tamura}, {Muto},
  {Kudo}, {Fukagawa}, {Fukue}, {Goto}, {Grady}, {Henning}, {Hodapp}, {Honda},
  {Inutsuka}, {Kokubo}, {Knapp}, {McElwain}, {Momose}, {Ohashi}, {Okamoto},
  {Takami}, {Turner}, {Wisniewski}, {Janson}, {Abe}, {Brandner}, {Carson},
  {Egner}, {Feldt}, {Golota}, {Guyon}, {Hayano}, {Hayashi}, {Hayashi}, {Ishii},
  {Kandori}, {Kusakabe}, {Matsuo}, {Mayama}, {Miyama}, {Morino}, {Moro-Martin},
  {Nishimura}, {Pyo}, {Suto}, {Suzuki}, {Takato}, {Terada}, {Thalmann},
  {Tomono}, {Watanabe}, {Yamada}, {Takami}, \& {Usuda}}]{hashimoto11}
{Hashimoto} J. {et~al.}, 2011, \apjl, 729, L17

\bibitem[{{Hayes} {et~al}\mbox{.}(2006){Hayes}, {Norman}, {Fiedler}, {Bordner},
  {Li}, {Clark}, {ud-Doula}, \& {Mac Low}}]{hayes06}
{Hayes} J.~C., {Norman} M.~L., {Fiedler} R.~A., {Bordner} J.~O., {Li} P.~S.,
  {Clark} S.~E., {ud-Doula} A., {Mac Low} M., 2006, \apjs, 165, 188

\bibitem[{{Heemskerk}, {Papaloizou} \& {Savonije}(1992){Heemskerk},
  {Papaloizou}, \& {Savonije}}]{heemskerk92}
{Heemskerk} M.~H.~M., {Papaloizou} J.~C., {Savonije} G.~J., 1992, \aap, 260,
  161

\bibitem[{{Hopkins}(2010)}]{hopkins10}
{Hopkins} P.~F., 2010, ArXiv e-prints

\bibitem[{{Inutsuka}, {Machida} \& {Matsumoto}(2010){Inutsuka}, {Machida}, \&
  {Matsumoto}}]{inutsuka10}
{Inutsuka} S.-i., {Machida} M.~N., {Matsumoto} T., 2010, \apjl, 718, L58

\bibitem[{{Isella} {et~al}\mbox{.}(2013){Isella}, {P{\'e}rez}, {Carpenter},
  {Ricci}, {Andrews}, \& {Rosenfeld}}]{isella13}
{Isella} A., {P{\'e}rez} L.~M., {Carpenter} J.~M., {Ricci} L., {Andrews} S.,
  {Rosenfeld} K., 2013, \apj, 775, 30

\bibitem[{{Juhasz} {et~al}\mbox{.}(2014){Juhasz}, {Benisty}, {Pohl},
  {Dullemond}, {Dominik}, \& {Paardekooper}}]{juhasz14}
{Juhasz} A., {Benisty} M., {Pohl} A., {Dullemond} C., {Dominik} C.,
  {Paardekooper} S.-J., 2014, ArXiv e-prints

\bibitem[{{Kimura} \& {Tsuribe}(2012)}]{kimura12}
{Kimura} S.~S., {Tsuribe} T., 2012, \pasj, 64, 116

\bibitem[{{Kratter} {et~al}\mbox{.}(2010){Kratter}, {Matzner}, {Krumholz}, \&
  {Klein}}]{kratter10b}
{Kratter} K.~M., {Matzner} C.~D., {Krumholz} M.~R., {Klein} R.~I., 2010, \apj,
  708, 1585

\bibitem[{{Kratter} \& {Murray-Clay}(2011)}]{kratter11}
{Kratter} K.~M., {Murray-Clay} R.~A., 2011, \apj, 740, 1

\bibitem[{{Landry} {et~al}\mbox{.}(2013){Landry}, {Dodson-Robinson}, {Turner},
  \& {Abram}}]{landry13}
{Landry} R., {Dodson-Robinson} S.~E., {Turner} N.~J., {Abram} G., 2013, \apj,
  771, 80

\bibitem[{{Laughlin} \& {Korchagin}(1996)}]{laughlin96}
{Laughlin} G., {Korchagin} V., 1996, \apj, 460, 855

\bibitem[{{Laughlin}, {Korchagin} \& {Adams}(1997){Laughlin}, {Korchagin}, \&
  {Adams}}]{laughlin97}
{Laughlin} G., {Korchagin} V., {Adams} F.~C., 1997, \apj, 477, 410

\bibitem[{{Laughlin}, {Korchagin} \& {Adams}(1998){Laughlin}, {Korchagin}, \&
  {Adams}}]{laughlin98}
{Laughlin} G., {Korchagin} V., {Adams} F.~C., 1998, \apj, 504, 945

\bibitem[{{Laughlin} \& {Rozyczka}(1996)}]{laughlin96b}
{Laughlin} G., {Rozyczka} M., 1996, \apj, 456, 279

\bibitem[{{Lin}, {Papaloizou} \& {Kley}(1993){Lin}, {Papaloizou}, \&
  {Kley}}]{lin93b}
{Lin} D.~N.~C., {Papaloizou} J.~C.~B., {Kley} W., 1993, \apj, 416, 689

\bibitem[{{Lin}(2012)}]{lin12b}
{Lin} M.-K., 2012, \mnras, 426, 3211

\bibitem[{{Lin}(2014)}]{lin14}
{Lin} M.-K., 2014, \mnras, 437, 575

\bibitem[{{Lin} \& {Cloutier}(2014)}]{lin14b}
{Lin} M.-K., {Cloutier} R., 2014, in IAU Symposium, Vol. 299, IAU Symposium,
  {Booth} M., {Matthews} B.~C., {Graham} J.~R., eds., pp. 218--219

\bibitem[{{Lin} \& {Papaloizou}(2011)}]{lin11b}
{Lin} M.-K., {Papaloizou} J.~C.~B., 2011, \mnras, 415, 1445

\bibitem[{{Lodato} \& {Rice}(2004)}]{lodato04}
{Lodato} G., {Rice} W.~K.~M., 2004, \mnras, 351, 630

\bibitem[{{Lodato} \& {Rice}(2005)}]{lodato05}
{Lodato} G., {Rice} W.~K.~M., 2005, \mnras, 358, 1489

\bibitem[{{Lynden-Bell} \& {Kalnajs}(1972)}]{lynden-bell72}
{Lynden-Bell} D., {Kalnajs} A.~J., 1972, \mnras, 157, 1

\bibitem[{{Mamatsashvili} \& {Rice}(2010)}]{mamat10}
{Mamatsashvili} G.~R., {Rice} W.~K.~M., 2010, \mnras, 406, 2050

\bibitem[{{Martin} {et~al}\mbox{.}(2012{\natexlab{a}}){Martin}, {Lubow},
  {Livio}, \& {Pringle}}]{martin12}
{Martin} R.~G., {Lubow} S.~H., {Livio} M., {Pringle} J.~E., 2012{\natexlab{a}},
  \mnras, 420, 3139

\bibitem[{{Martin} {et~al}\mbox{.}(2012{\natexlab{b}}){Martin}, {Lubow},
  {Livio}, \& {Pringle}}]{martin12b}
{Martin} R.~G., {Lubow} S.~H., {Livio} M., {Pringle} J.~E., 2012{\natexlab{b}},
  \mnras, 423, 2718

\bibitem[{{Masset}(2000{\natexlab{a}})}]{masset00a}
{Masset} F., 2000{\natexlab{a}}, \aaps, 141, 165

\bibitem[{{Masset}(2000{\natexlab{b}})}]{masset00b}
{Masset} F.~S., 2000{\natexlab{b}}, in Astronomical Society of the Pacific
  Conference Series, Vol. 219, Disks, Planetesimals, and Planets, {Garz{\'o}n}
  G., {Eiroa} C., {de Winter} D., {Mahoney} T.~J., eds., pp. 75--+

\bibitem[{{Matzner} \& {Levin}(2005)}]{matzner05}
{Matzner} C.~D., {Levin} Y., 2005, \apj, 628, 817

\bibitem[{{Mayer} {et~al}\mbox{.}(2004){Mayer}, {Quinn}, {Wadsley}, \&
  {Stadel}}]{mayer04}
{Mayer} L., {Quinn} T., {Wadsley} J., {Stadel} J., 2004, \apj, 609, 1045

\bibitem[{{Michael} \& {Durisen}(2010)}]{michael10}
{Michael} S., {Durisen} R.~H., 2010, \mnras, 406, 279

\bibitem[{{Mignone} {et~al}\mbox{.}(2007){Mignone}, {Bodo}, {Massaglia},
  {Matsakos}, {Tesileanu}, {Zanni}, \& {Ferrari}}]{mignone07}
{Mignone} A., {Bodo} G., {Massaglia} S., {Matsakos} T., {Tesileanu} O., {Zanni}
  C., {Ferrari} A., 2007, \apjs, 170, 228

\bibitem[{{M{\"u}ller}, {Kley} \& {Meru}(2012){M{\"u}ller}, {Kley}, \&
  {Meru}}]{muller12}
{M{\"u}ller} T.~W.~A., {Kley} W., {Meru} F., 2012, \aap, 541, A123

\bibitem[{{Muto} {et~al}\mbox{.}(2012){Muto}, {Grady}, {Hashimoto}, {Fukagawa},
  {Hornbeck}, {Sitko}, {Russell}, {Werren}, {Cur{\'e}}, {Currie}, {Ohashi},
  {Okamoto}, {Momose}, {Honda}, {Inutsuka}, {Takeuchi}, {Dong}, {Abe},
  {Brandner}, {Brandt}, {Carson}, {Egner}, {Feldt}, {Fukue}, {Goto}, {Guyon},
  {Hayano}, {Hayashi}, {Hayashi}, {Henning}, {Hodapp}, {Ishii}, {Iye},
  {Janson}, {Kandori}, {Knapp}, {Kudo}, {Kusakabe}, {Kuzuhara}, {Matsuo},
  {Mayama}, {McElwain}, {Miyama}, {Morino}, {Moro-Martin}, {Nishimura}, {Pyo},
  {Serabyn}, {Suto}, {Suzuki}, {Takami}, {Takato}, {Terada}, {Thalmann},
  {Tomono}, {Turner}, {Watanabe}, {Wisniewski}, {Yamada}, {Takami}, {Usuda}, \&
  {Tamura}}]{muto12}
{Muto} T. {et~al.}, 2012, \apjl, 748, L22

\bibitem[{{Narayan}, {Goldreich} \& {Goodman}(1987){Narayan}, {Goldreich}, \&
  {Goodman}}]{narayan87}
{Narayan} R., {Goldreich} P., {Goodman} J., 1987, \mnras, 228, 1

\bibitem[{{Nelson} {et~al}\mbox{.}(1998){Nelson}, {Benz}, {Adams}, \&
  {Arnett}}]{nelson98}
{Nelson} A.~F., {Benz} W., {Adams} F.~C., {Arnett} D., 1998, \apj, 502, 342

\bibitem[{{Ogilvie}(2008)}]{ogilvie08}
{Ogilvie} G.~I., 2008, \mnras, 388, 1372

\bibitem[{{Okazaki}(1997)}]{okasaki97}
{Okazaki} A.~T., 1997, \aap, 318, 548

\bibitem[{{Paardekooper}(2012)}]{paardekooper12}
{Paardekooper} S.-J., 2012, \mnras, 421, 3286

\bibitem[{{Papaloizou} \& {Savonije}(1991)}]{papaloizou91}
{Papaloizou} J.~C., {Savonije} G.~J., 1991, \mnras, 248, 353

\bibitem[{{Papaloizou}(2002)}]{papaloizou02}
{Papaloizou} J.~C.~B., 2002, \aap, 388, 615

\bibitem[{{Papaloizou} {et~al}\mbox{.}(2007){Papaloizou}, {Nelson}, {Kley},
  {Masset}, \& {Artymowicz}}]{papaloizou07}
{Papaloizou} J.~C.~B., {Nelson} R.~P., {Kley} W., {Masset} F.~S., {Artymowicz}
  P., 2007, in Protostars and Planets V, {Reipurth} B., {Jewitt} D., {Keil} K.,
  eds., pp. 655--668

\bibitem[{{Papaloizou} \& {Pringle}(1985)}]{papaloizou85}
{Papaloizou} J.~C.~B., {Pringle} J.~E., 1985, \mnras, 213, 799

\bibitem[{{Papaloizou} \& {Savonije}(2006)}]{papaloizou06c}
{Papaloizou} J.~C.~B., {Savonije} G.~J., 2006, \aap, 456, 1097

\bibitem[{{P{\'e}rez} {et~al}\mbox{.}(2014){P{\'e}rez}, {Isella}, {Carpenter},
  \& {Chandler}}]{perez14}
{P{\'e}rez} L.~M., {Isella} A., {Carpenter} J.~M., {Chandler} C.~J., 2014,
  \apjl, 783, L13

\bibitem[{{Rafikov}(2005)}]{rafikov05}
{Rafikov} R.~R., 2005, \apjl, 621, L69

\bibitem[{{Rafikov}(2009)}]{rafikov09}
{Rafikov} R.~R., 2009, \apj, 704, 281

\bibitem[{{Rice} \& {Armitage}(2009)}]{rice09}
{Rice} W.~K.~M., {Armitage} P.~J., 2009, \mnras, 396, 2228

\bibitem[{{Rice} {et~al}\mbox{.}(2011){Rice}, {Armitage}, {Mamatsashvili},
  {Lodato}, \& {Clarke}}]{rice11}
{Rice} W.~K.~M., {Armitage} P.~J., {Mamatsashvili} G.~R., {Lodato} G., {Clarke}
  C.~J., 2011, \mnras, 418, 1356

\bibitem[{{Rice}, {Lodato} \& {Armitage}(2005){Rice}, {Lodato}, \&
  {Armitage}}]{rice05}
{Rice} W.~K.~M., {Lodato} G., {Armitage} P.~J., 2005, \mnras, 364, L56

\bibitem[{{Rivinius}, {Carciofi} \& {Martayan}(2013){Rivinius}, {Carciofi}, \&
  {Martayan}}]{rivinius13}
{Rivinius} T., {Carciofi} A.~C., {Martayan} C., 2013, \aapr, 21, 69

\bibitem[{{Rogers} \& {Wadsley}(2012)}]{rogers12}
{Rogers} P.~D., {Wadsley} J., 2012, \mnras, 423, 1896

\bibitem[{{Ryu} \& {Goodman}(1992)}]{ryu92}
{Ryu} D., {Goodman} J., 1992, \apj, 388, 438

\bibitem[{Shu(1991)}]{shu91}
Shu F., 1991, The Physics of Astrophysics: Gas dynamics, Series of books in
  astronomy. University Science Books

\bibitem[{{Shu} {et~al}\mbox{.}(1990){Shu}, {Tremaine}, {Adams}, \&
  {Ruden}}]{shu90}
{Shu} F.~H., {Tremaine} S., {Adams} F.~C., {Ruden} S.~P., 1990, \apj, 358, 495

\bibitem[{{Stamatellos} \& {Whitworth}(2008)}]{stamatellos08}
{Stamatellos} D., {Whitworth} A.~P., 2008, \aap, 480, 879

\bibitem[{{Toomre}(1964)}]{toomre64}
{Toomre} A., 1964, \apj, 139, 1217

\bibitem[{{Tremaine}(2001)}]{tremaine01}
{Tremaine} S., 2001, \aj, 121, 1776

\bibitem[{{Tsukamoto}, {Machida} \& {Inutsuka}(2013){Tsukamoto}, {Machida}, \&
  {Inutsuka}}]{tsukamoto13}
{Tsukamoto} Y., {Machida} M.~N., {Inutsuka} S.-i., 2013, \mnras, 436, 1667

\bibitem[{{Tsukamoto} {et~al}\mbox{.}(2015){Tsukamoto}, {Takahashi}, {Machida},
  \& {Inutsuka}}]{tsukamoto15}
{Tsukamoto} Y., {Takahashi} S.~Z., {Machida} M.~N., {Inutsuka} S., 2015,
  \mnras, 446, 1175

\bibitem[{{Turner} \& {Sano}(2008)}]{turner08}
{Turner} N.~J., {Sano} T., 2008, \apjl, 679, L131

\bibitem[{{van der Marel} {et~al}\mbox{.}(2013){van der Marel}, {van Dishoeck},
  {Bruderer}, {Birnstiel}, {Pinilla}, {Dullemond}, {van Kempen}, {Schmalzl},
  {Brown}, {Herczeg}, {Mathews}, \& {Geers}}]{marel13}
{van der Marel} N. {et~al.}, 2013, Science, 340, 1199

\bibitem[{{van der Plas} {et~al}\mbox{.}(2014){van der Plas}, {Casassus},
  {M{\'e}nard}, {Perez}, {Thi}, {Pinte}, \& {Christiaens}}]{plas14}
{van der Plas} G., {Casassus} S., {M{\'e}nard} F., {Perez} S., {Thi} W.~F.,
  {Pinte} C., {Christiaens} V., 2014, \apjl, 792, L25

\bibitem[{{Zhu}, {Hartmann} \& {Gammie}(2010){Zhu}, {Hartmann}, \&
  {Gammie}}]{zhu10b}
{Zhu} Z., {Hartmann} L., {Gammie} C., 2010, \apj, 713, 1143

\bibitem[{{Zhu} {et~al}\mbox{.}(2009){Zhu}, {Hartmann}, {Gammie}, \&
  {McKinney}}]{zhu09}
{Zhu} Z., {Hartmann} L., {Gammie} C., {McKinney} J.~C., 2009, \apj, 701, 620

\bibitem[{{Zhu} {et~al}\mbox{.}(2010){Zhu}, {Hartmann}, {Gammie}, {Book},
  {Simon}, \& {Engelhard}}]{zhu10}
{Zhu} Z., {Hartmann} L., {Gammie} C.~F., {Book} L.~G., {Simon} J.~B.,
  {Engelhard} E., 2010, \apj, 713, 1134

\bibitem[{{Zhu} {et~al}\mbox{.}(2012){Zhu}, {Hartmann}, {Nelson}, \&
  {Gammie}}]{zhu12}
{Zhu} Z., {Hartmann} L., {Nelson} R.~P., {Gammie} C.~F., 2012, \apj, 746, 110

\end{thebibliography}

\appendix
\section{The background torque density in a three-dimensional disc with
  a fixed temperature profile}\label{tbg_deriv}
We give a brief derivation of the angular momentum exchange between
linear perturbations and the background disc. We consider a
three-dimensional disc in which the equilibrium pressure and density
are related by 
\begin{align}\label{iso_cond}
  p = c_s^2(R,z)F(\rho),
\end{align} 
where $F(\rho)$ is an arbitrary function of $\rho$ with dimensions of
mass per unit volume, and $c_s$ is a prescribed function of
position with dimensions of velocity squared. The equilibrium disc
satisfies 
\begin{align}
  R\Omega^2(R,z) &= \frac{1}{\rho}\frac{\p p}{d R} +
  \frac{\p\Phi_\mathrm{tot}}{\p R}, \\
  0 &= \frac{1}{\rho}\frac{\p p}{\p z} + \frac{\p\Phi_\mathrm{tot}}{\p
    z}. 
\end{align}
Note that, in general, the equilibrium rotation $\Omega$ depends on
$R$ and $z$. 

We begin with the linearised equation of motion in terms of the
Lagrangian displacement $\bm{\xi}$ as given by \cite{lin93b} but with an
additional potential perturbation, 
\begin{align}\label{lagragian_pert}
  &\frac{D^2\bm{\xi}}{Dt^2} +
  2\Omega\hat{\bm{z}}\times\frac{D\bm{\xi}}{Dt}  \notag \\ &= -
  \frac{\nabla \delta p }{\rho} + \frac{\delta\rho}{\rho^2}\nabla p  
  -\nabla\delta\Phi_d - R
  \hat{\bm{R}}\left(\bm{\xi}\cdot\nabla\Omega^2\right) \notag \\
  & = -\nabla\left(\frac{\delta p}{\rho} + \delta \Phi_d\right) -
  \frac{\delta p}{\rho}\frac{\nabla\rho}{\rho} +
  \frac{\delta\rho}{\rho}\frac{\nabla p}{\rho} -  R
  \hat{\bm{R}}\left(\bm{\xi}\cdot\nabla\Omega^2\right),
\end{align}
where $D/Dt \equiv \p_t + \ii m \Omega$ for perturbations with
azimuthal dependence in the form $\exp\left(\ii m \phi\right)$, and  
the $\delta$ quantities denote Eulerian perturbations. 

As explained in \cite{lin11b}, a conservation law for the angular
momentum of the perturbation may be obtained by taking the dot product
between Eq. \ref{lagragian_pert} and $(-m/2)\rho\bm{\xi}^*$, then
taking the imaginary part afterwards. The left hand side becomes the
 rate of change of angular momentum density. The first term on the right hand side (RHS)
becomes 
\begin{align}\label{angflux1}
  &-\frac{m}{2}\imag\left[-\rho\bm{\xi}^*\cdot\nabla\left(\frac{\delta p}{\rho} + \delta
    \Phi_d\right)\right] \notag\\ 
&= \frac{m}{2}\imag\left\{\nabla\cdot\left[\rho\bm{\xi}^*\left(\frac{\delta p}{\rho} + \delta
    \Phi_d \right) + \frac{1}{4\pi
    G}\delta\Phi_d\nabla\delta\Phi_d^*\right]\right\} \notag\\
&+ \frac{m}{2}\imag\left(\delta\rho^*\frac{\delta p}{\rho}\right),
\end{align}
where $\delta\rho = - \nabla\cdot\left(\rho\bm{\xi}\right)$ and
$\nabla^2\delta\Phi_d = 4\pi G \delta \rho$ have been used. The 
terms in square brackets on the RHS of Eq. \ref{angflux1} is (minus) the 
angular momentum flux. The second term on RHS of Eq. \ref{angflux1}, 
together with the remaining terms on the RHS of 
Eq. \ref{lagragian_pert} constitutes the background torque. That is, 
\begin{align}\label{tbg_3d}
  T_\mathrm{BG} = \frac{m}{2}\imag\left[
    \frac{\delta p}{\rho} \Delta\rho^* -
    \frac{\delta\rho}{\rho}\bm{\xi}^*\cdot\nabla p  
    + \rho \xi_R^*\xi_z\frac{\p\left(R\Omega^2\right)}{\p z}\right],
\end{align}
where $\Delta\rho = \delta\rho + \bm{\xi}\cdot\nabla\rho$ is the
Lagrangian density perturbation. 

So far we have not invoked an energy equation. For 
adiabatic perturbations $T_\mathrm{BG}$ is zero, and we recover the
same statement of angular momentum conservation as in \cite{lin93b} but 
modified by self-gravity in the fluxes. 

However, if we impose the equilibrium relation
Eq. \ref{iso_cond} to hold in the perturbed state, then
\begin{align}
  \delta p = c_s^2(R,z) F^\prime \delta\rho,
\end{align}
where $F^\prime = dF/d\rho$. Inserting this into Eq. \ref{tbg_3d}, we
obtain
\begin{align}\label{tbg_3d_2}
  T_\mathrm{BG} = -\frac{m}{2}\frac{p}{\rho
    c_s^2}\imag\left[\delta\rho\bm{\xi}^*\cdot\nabla c_s^2 +
    \xi_R^*\xi_z \left(\frac{\p\rho}{\p z}\frac{\p c_s^2}{\p R} -
      \frac{\p\rho}{\p R}\frac{\p c_s^2}{\p z}\right)\right],
\end{align}
where the equilibrium equations were used. 
At this point setting $\xi_z=0$ gives $T_\mathrm{BG}$ for
perturbations with no vertical motion, 
\begin{align}
  T_\mathrm{BG,2D} = -\frac{m}{2}\frac{p}{\rho
    c_s^2}\imag\left(\delta\rho \xi_R^*  \p_R c_s^2 \right),
\end{align}
and is equivalent to the 2D
expression, Eq. \ref{baroclinic_torque}, with $\delta\rho $ replaced
by $\delta \Sigma$ and $p=c_s^2\rho$. 

In fact, we can simplify Eq. \ref{tbg_3d_2} in the general case by
using $\delta\rho = - \rho\nabla\cdot\bm{\xi} -
\bm{\xi}\cdot\nabla\rho$, giving
\begin{align}\label{tbg_general}
  T_\mathrm{BG} = \frac{m}{2}\frac{p}{\rho
    c_s^2}\imag\left[\rho\left(\nabla\cdot\bm{\xi}\right)\bm{\xi}^*\cdot\nabla
  c_s^2\right].  
\end{align} 
For a barotropic fluid $p=p(\rho)$, the function $c_s^2$ can be
taken as constant (Eq. \ref{iso_cond}) for which $T_\mathrm{BG}$
vanishes. When there is a forced temperature gradient,
Eq. \ref{tbg_general} indicates a torque is applied to compressible
perturbations ($\nabla\cdot\bm{\xi}\neq0$) if there is motion along
the temperature gradient ($\bm{\xi}\cdot\nabla c_s^2 \neq 0$).   




\section{Relation between horizontal Lagrangian displacements for
  local, low frequency disturbances}\label{horizontal_displacements}
Here, we aim to relate the horizontal Lagrangian displacements $\xi_R$
and $\xi_\phi$ in the local approximation. Using the local solution to
the Poisson equation 
\begin{align}
  \delta \Phi_m = -\frac{2\pi G}{|k|} \delta\Sigma_m 
\end{align}
\citep{shu91}, the linearised azimuthal equation of motion becomes 
\begin{align} 
  - \ii\sbar \delta v_{\phi m}  + \frac{\kappa^2}{2\Omega}\delta v_{Rm} = -\frac{\ii
    m}{R\Sigma}\left(c_s^2 - \frac{2\pi G
      \Sigma}{|k|}\right)\delta\Sigma_m. 
\end{align}
Next, we replace the surface density perturbation
$\delta \Sigma_m = -\ii k \Sigma \xi_R$, and use the expressions
\begin{align}
  &\delta v_{Rm} = -\ii\sbar\xi_R,\\
  &\delta v_{\phi m} = -\ii\sbar\xi_\phi - \frac{\ii R
    \p_R\Omega}{\sbar} \delta v_{Rm}
\end{align}
\citep{papaloizou85} to obtain 
\begin{align}
  -\sbar^2\xi_\phi - 2\ii\sbar\Omega \xi_R =
  \frac{m}{kR}\left(\kappa^2 - \sbar^2\right)\xi_R, 
\end{align}
where the dispersion relation Eq. \ref{dispersion} was used. 
In the local approximation, $|kR|\gg m$ by assumption. 
 Hence the RHS of this equation can be neglected. Then 
\begin{align}
  \xi_\phi \simeq -\frac{2\ii\Omega}{\sbar}\xi_R.
\end{align} 
For low-frequency modes we have $\sbar \simeq -m\Omega$, so
$\xi_\phi\simeq 2\ii\xi_R/m$, as used in the main text.

\section{The confined spiral as an external potential}\label{disc-planet}
Let us treat the one-armed spiral confined in $R\in[R_1,R_2]$ as an  
external potential of the form $\Phi_\mathrm{ext}(R)\cos{\left(\phi -
    \Omega_pt\right)}$. We take 
\begin{align} 
  \Phi_\mathrm{ext}
  =-\frac{GM_\mathrm{ring}}{\overline{R}}b^{1}_{1/2}(\beta),   
\end{align}
where $M_\mathrm{ring}$ is the disc mass contained within
$R\in[R_1,R_2]$, $\overline{R} = (R_1+R_2)/2$ is the approximate radial
location of the spiral, $b_{n}^m(\beta)$ is the Laplace coefficient
and $\beta = R/\overline{R}$. This form of
$\Phi_\mathrm{ext}$ is the $m=1$ component of the gravitational
potential of an external satellite on a circular orbit
\citep{goldreich79}.    


We expect the external potential to exert a torque on the disc at the
Lindblad and co-rotation resonances. At the outer Lindblad resonance
(OLR), this torque is 
\begin{align}
  \Gamma_L =
  \frac{\pi^2\Sigma_L}{3\Omega_L\Omega_p}
  \left[\left.R_L\frac{d\Phi_\mathrm{ext}}{dR}\right|_L + 2\left(1 -
      \frac{\Omega_p}{\Omega_L}\right)\Phi_\mathrm{ext}\right]^2, 
\end{align}   
where a Keplerian disc has been assumed and subscript $L$ denotes
evaluation at the OLR, $R=R_L$. (The inner Lindblad resonance does not
exist for the pattern speeds observed in our simulations.) 

If we associate the external potential with  an angular momentum magnitude of  
$J_\mathrm{ext}  = M_\mathrm{ring}\overline{R}^2\Omega_p$, we can
calculate a rate of change of angular momentum $\gamma_L=\Gamma_L/J_\mathrm{ext}$. Then  
\begin{align}
  \frac{\gamma_L}{\Omega_p} = &\frac{\pi
    h}{3Q_L}\left(\frac{M_p}{M_*}\right)\left(\frac{R_L}{\overline{R}}\right)\left(\frac{R_c}{\overline{R}}\right) 
  ^3\left(\frac{R_L}{R_c}\right)^{-3/2}\notag\\ 
  &\times
  \left\{\frac{R_L}{\overline{R}}\left.\frac{db_{1/2}^1}{d\beta}\right|_L
    + 2\left[1 - 
      \left(\frac{R_c}{R_L}\right)^{-3/2}\right]b_{1/2}^1(\beta_L)\right\}^2. 
\end{align}
Inserting $h=0.05$, $Q_L=10$, $M_\mathrm{ring} = 0.05M_*$,
$R_L=7.2R_0$, $R_c=4.4R_0$ and $\overline{R}=1.5R_0$ from our fiducial
FARGO simulation, we get
\begin{align}
  \gamma_L \sim 5\times10^{-4}\Omega_p. 
\end{align}
For the co-rotation torque, we use the result
\begin{align}
  \Gamma_c = \left.
    \pi^2\Phi_\mathrm{ext}^2\left(\frac{d\Omega}{dR}\right)^{-1}\frac{d}{dR}\left(\frac{2\Sigma}{\Omega}\right)\right|_{c}     
\end{align}
for a Keplerian disc, where subscript $c$ denotes evaluation at
co-rotation radius $R=R_c$. For a power-law surface density profile
$\Sigma\propto R^{-s}$ we have
\begin{align}
\frac{\gamma_c}{\Omega_p} = \frac{4}{3}\frac{\pi h}{Q_c}
\left(\frac{M_\mathrm{ring}}{M_*}\right)\left(\frac{R_c}{\overline{R}}\right)^4\left(s-\frac{3}{2}\right) 
\left[b_{1/2}^1(\beta_c)\right]^2    
\end{align}
Using the above parameter values with $s=2$ and $Q_c=10$, we obtain a
rate
\begin{align}
  \gamma_c\sim 6\times 10^{-4}\Omega_p. 
\end{align}

The torque exerted on the disc at the OLR by an external potential is
positive, while that at co-rotation depends 
on the gradient of potential vorticity there \citep{goldreich79}. For
our disc models with surface density $\Sigma\propto R^{-2}$ in the
outer disc, this co-rotation torque is positive. 
This means that the one-armed spiral loses
angular momentum by launching density waves with positive angular
momentum at the OLR, and by applying a 
positive co-rotation torque on the disc. In principle, this
interaction is destabilising because the one-armed spiral has 
negative angular momentum \citep{lin11b}. 

 However, the above estimates for $\gamma_L$ and $\gamma_c$ are much smaller than that due to the 
imposed temperature gradient as measured in the simulations 
($\gamma\sim∼0.1\Omega_p$). We conclude that for our disc models, 
the Lindblad and co-rotation resonances have negligible effects
on the growth of the one-armed spiral in the inner disc (but it could be  important 
in other parameter regimes).

\end{document}